\documentclass[modern]{aastex62}

\usepackage{url}
\usepackage{amsmath}
\usepackage{mathtools}
\usepackage{amssymb}
\usepackage{natbib}
\usepackage{graphicx}
\usepackage{calc}
\usepackage{etoolbox}
\usepackage{xspace}
\usepackage[T1]{fontenc} 
\usepackage{textcomp}
\usepackage{ifxetex}
\ifxetex
  \usepackage{fontspec}
  \defaultfontfeatures{Extension = .otf}
\fi
\usepackage{fontawesome}
\usepackage{listings}
\usepackage{nicefrac}
\usepackage{booktabs}
\usepackage{longtable}

\renewcommand{\eqref}[1]{\ref{eq:#1}}

\definecolor{linkcolor}{rgb}{0.1216,0.4667,0.7059}
\definecolor{testpasscolor}{rgb}{0.13333333,0.5254902,0.22745098}
\definecolor{testmissingcolor}{rgb}{1.0,0.88,0.30}
\definecolor{testfailcolor}{rgb}{0.79607843,0.14117647,0.19215686}
\newcommand{\codeicon}{{\color{linkcolor}\faCloudDownload}}
\newcommand{\testmissingicon}{{\color{testmissingcolor}\faQuestion}}
\newcommand{\testpassicon}{{\color{testpasscolor}\faCheck}}
\newcommand{\testfailicon}{{\color{testfailcolor}\faTimes}}
\newcommand{\codelink}[1]{\href{https://github.com/rodluger/mapping_stellar_surfaces/blob/paper1-arxiv/paper1/figures/#1.py}{\codeicon}\,\,}
\newcommand{\animlink}[1]{\href{https://github.com/rodluger/mapping_stellar_surfaces/blob/paper1-arxiv/paper1/figures/#1.gif}{\animicon}\,\,}

\newtagform{eqtag}[]{(}{)}
\newcommand{\currentlabel}{None}
\newenvironment{proof}[1]{%
  \ifstrempty{#1}{%
    \renewtagform{eqtag}[]{\raisebox{-0.1em}{{\testmissingicon}}\,(}{)}%
  }{%
    \renewtagform{eqtag}[]{\href{https://github.com/rodluger/mapping_stellar_surfaces/blob/paper1-arxiv/paper1/tests/#1.py}{\raisebox{-0.1em}{\input{tests/#1.tex}}}\,(}{)}%
  }%
  \usetagform{eqtag}%
  \renewcommand{\currentlabel}{#1}
  \align%
}{%
  \endalign%
  \renewtagform{eqtag}[]{(}{)}%
  \usetagform{eqtag}%
  \message{<<<\currentlabel: \theequation>>>}%
}

\newcommand{\oscaption}[2]{\caption{#2 \codelink{#1}}}

\definecolor{codegreen}{rgb}{0,0.6,0}
\definecolor{codegray}{rgb}{0.5,0.5,0.5}
\definecolor{codepurple}{rgb}{0.58,0,0.82}
\definecolor{backcolour}{rgb}{0.95,0.95,0.95}
\lstdefinestyle{mystyle}{
  backgroundcolor=\color{backcolour},
  commentstyle=\color{codegreen},
  keywordstyle=\color{magenta},
  numberstyle=\tiny\color{codegray},
  stringstyle=\color{codepurple},
  basicstyle=\small\ttfamily,
  breakatwhitespace=false,
  breaklines=true,
  captionpos=b,
  keepspaces=true,
  numbers=left,
  numbersep=5pt,
  showspaces=false,
  showstringspaces=false,
  showtabs=false,
  tabsize=2,
  aboveskip=1em,
  belowskip=1em,
  keywords=[2]{map},
  keywordstyle=[2]{\color{black!80!black}},
  upquote=true
}
\renewcommand\quad{\hskip\fontdimen3\font}

\usepackage{stackengine,scalerel}
\def\lnlam{\ThisStyle{\ensurestackMath{\stackon[-2.4\LMpt]{%
        \SavedStyle\lambda}{\kern-.5pt\kern\LMpt\rule{1\LMex}{.25pt+.15\LMpt}}}}}

\usepackage{xifthen}
\usepackage{stackengine}
\usepackage{tabstackengine}
\usepackage{array}
\usepackage{upgreek}
\usepackage[bbgreekl]{mathbbol}
\usepackage{afterpage}
\usepackage[bb=boondox]{mathalpha}

\newcommand{\starry}{\textsf{starry}\xspace}

\newcommand{\Python}{\textsf{Python}\xspace}

\makeatletter
\def\Ddots{\mathinner{\mkern1mu\raise\p@
        \vbox{\kern7\p@\hbox{.}}\mkern2mu
        \raise4\p@\hbox{.}\mkern2mu\raise7\p@\hbox{.}\mkern1mu}}
\makeatother

\DeclareFontFamily{U}{mathc}{}
\DeclareFontShape{U}{mathc}{m}{it}{<->s*[1.03] mathc10}{}
\DeclareMathAlphabet{\mathscr}{U}{mathc}{m}{it}

\usepackage{etoolbox}
\makeatletter 
\patchcmd{\env@cases}{\quad}{\qquad\qquad}{}{}
\makeatother

\defcitealias{PaperII}{Paper II}

\usepackage{enumitem}

\newcommand{\shrinkage}{{variance reduction\,}}

\begin{document}

\title{%
    \vspace{-3em}
    \textbf{
        Mapping stellar surfaces\\
        I: Degeneracies in the rotational light curve problem
    }
}

\author[0000-0002-0296-3826]{Rodrigo Luger}\altaffiliation{Flatiron Fellow}
\email{rluger@flatironinstitute.org}
\affil{Center~for~Computational~Astrophysics,~Flatiron~Institute,~New~York,~NY}
\affil{Virtual~Planetary~Laboratory, University~of~Washington, Seattle, WA}
\author[0000-0002-9328-5652]{Daniel Foreman-Mackey}
\affil{Center~for~Computational~Astrophysics,~Flatiron~Institute,~New~York,~NY}
\author[0000-0002-3385-8391]{Christina Hedges}
\affil{Bay~Area~Environmental~Research~Institute,~Moffett~Field,~CA}
\affil{NASA~Ames~Research~Center,~Moffett~Field,~CA}
\author[0000-0003-2866-9403]{David W. Hogg}
\affil{Center~for~Computational~Astrophysics,~Flatiron~Institute,~New~York,~NY}
\affil{Center~for~Cosmology~and~Particle~Physics,~New~York~University,~New~York,~NY}
\affil{Center~for~Data~Science,~New~York~University,~New~York,~NY}
\affil{Max-Planck-Institut~f\"ur~Astronomie,~Heidelberg,~Germany}


\features{open-source figures \codeicon; equation unit tests: 5 passed \testpassicon, 0 failed \testfailicon
}

\begin{abstract}
    Thanks to missions like \emph{Kepler} and \emph{TESS}, we now
    have access to tens of thousands of high precision, fast
    cadence, and long baseline stellar photometric observations.
    In principle, these light curves encode a vast amount of information about
    stellar variability and, in particular, about the distribution of
    starspots and other features on their surfaces.
    Unfortunately, the problem of inferring stellar surface properties
    from a rotational light curve is famously ill-posed,
    as it often does not admit a unique solution.
    Inference about the number, size, contrast, and location of spots
    can therefore depend very strongly on the assumptions of the model,
    the regularization scheme, or the prior.
    The goal of this paper is twofold:
    (1) to explore the various degeneracies affecting the stellar
    light curve ``inversion'' problem and their effect on
    what can and cannot be learned from a stellar surface
    given unresolved photometric measurements; and
    (2) to motivate ensemble analyses of the light curves of
    many stars at once as a powerful data-driven alternative to common
    priors adopted in the literature.
    We further derive novel results on the dependence of the null space
    on stellar inclination and limb darkening and
    show that single-band photometric measurements cannot uniquely
    constrain quantities like the total
    spot coverage without the use of strong priors.
    This is the first in a series of papers devoted to the development
    of novel algorithms and tools for the analysis of
    stellar light curves and spectral time series, with the
    explicit goal of enabling statistically robust inference
    about their surface properties.
\end{abstract}

\section{Introduction}
\label{sec:intro}

The advent of space-based precision photometry with missions such as
\emph{Kepler} \citep{Borucki2010} and \emph{TESS} \citep{Ricker2015}
has led to a renewed interest in the modeling of stellar light curves,
and, in particular, in understanding what these light curves can tell
us about the surfaces of stars across the HR diagram. One of the dominant
sources of stellar light curve variability is the modulation caused
by starspots rotating in and out of view.
Dark spots arise due to the suppression of convection in regions of
intense magnetic field, resulting in a locally cooler (and hence darker)
photosphere. Bright spots can similarly
arise in the photosphere as faculae or in the chromosphere as plages, and
are also magnetically driven \citep[e.g.,][]{Berdyugina2005}.
Constraining the sizes,
contrasts, locations, and number of spots on stars can therefore reveal
information about stellar magnetic activity, stellar interior structure,
and how these quantities vary across spectral type and over time
\citep[e.g.,][]{Garraffo2018}.
A detailed understanding of the stellar surface is also crucial to
mitigating systematics in the radial velocity search for exoplanets
\citep[e.g.,][]{Lanza2011} and in the spectroscopic characterization of their
atmospheres \citep[e.g.,][]{Rackham2018}.

\begin{figure}[t!]
    \begin{centering}
        \includegraphics[width=\linewidth]{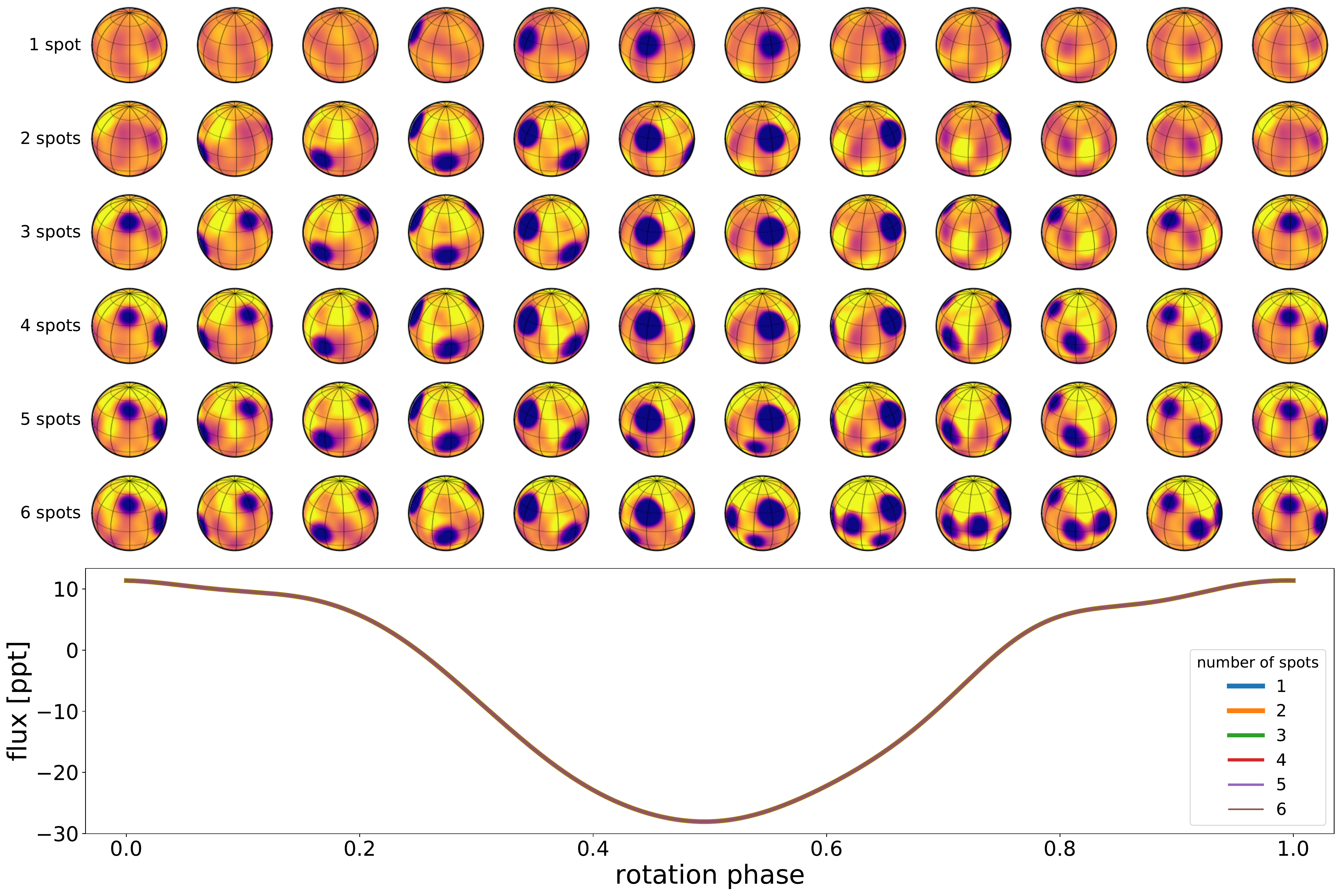}
        \oscaption{degeneracies}{%
            The fundamental limitations of the mapping problem. Each row
            corresponds to a stellar surface with a different number of
            dark spots seen at various phases at an inclination $I=60^\circ$;
            all images are shown on the same color scale.
            The bottom panel shows the light curves of each of these stars.
            All six light curves are indistinguishable from each other, even
            at infinite signal to noise. See text for details.
            \label{fig:degeneracies}
        }
    \end{centering}
\end{figure}

To date, most studies aimed at inferring stellar surface
properties from light curves follow one of two broad approaches. The
first is to model the stellar surface as a collection of one or more
discrete, circular, uniform contrast dark or bright spots on a uniform
intensity photosphere.
The advantage of this approach is that
the light curve can be computed efficiently and in some cases even
analytically \citep[e.g.,][]{Davenport2015,Morris2017,Morris2020b}.
The second approach is to discretize the surface at some resolution
and compute the emergent flux as a weighted sum of the visible pixel intensities.
This approach is more flexible, since it is not limited to surfaces
composed of distinct circular spots \citep[e.g.,][]{Harmon2000,Roettenbacher2017}.
Both approaches rely on an explicit \emph{forward model}, a prescription for
how to generate data given a set of parameters. In most cases, however, we are interested
in the inverse problem: constraining the parameters given data.
Unfortunately, the inverse problem is not only difficult---as it requires
a large number of forward model evaluations to find the parameter values
that are most consistent with the
data---but also formally \emph{ill-posed}.
While the mapping from a stellar surface to a light curve (the forward
problem) is unique, the mapping from a light curve to a surface
(the inverse problem) is not:
given any light curve, there exist an infinite number of surfaces that
could have generated it. These degeneracies are illustrated in
Figure~\ref{fig:degeneracies}, where six synthetic stellar surfaces
are shown (rows) at twelve different phases (columns), all at an
inclination $I=60^\circ$.%
\footnote{
    In this paper, we adopt the common convention where $I$ is the angle
    between the stellar spin axis and the line of sight,
    spanning $-90^\circ \leq I \leq 90^\circ$. An inclination $I=0^\circ$ therefore
    corresponds to a pole-on orientation.
}
Each surface
consists of a different number of dark spots on a brighter, heterogeneous
background.

\begin{figure}[t!]
    \begin{centering}
        \includegraphics[width=\linewidth]{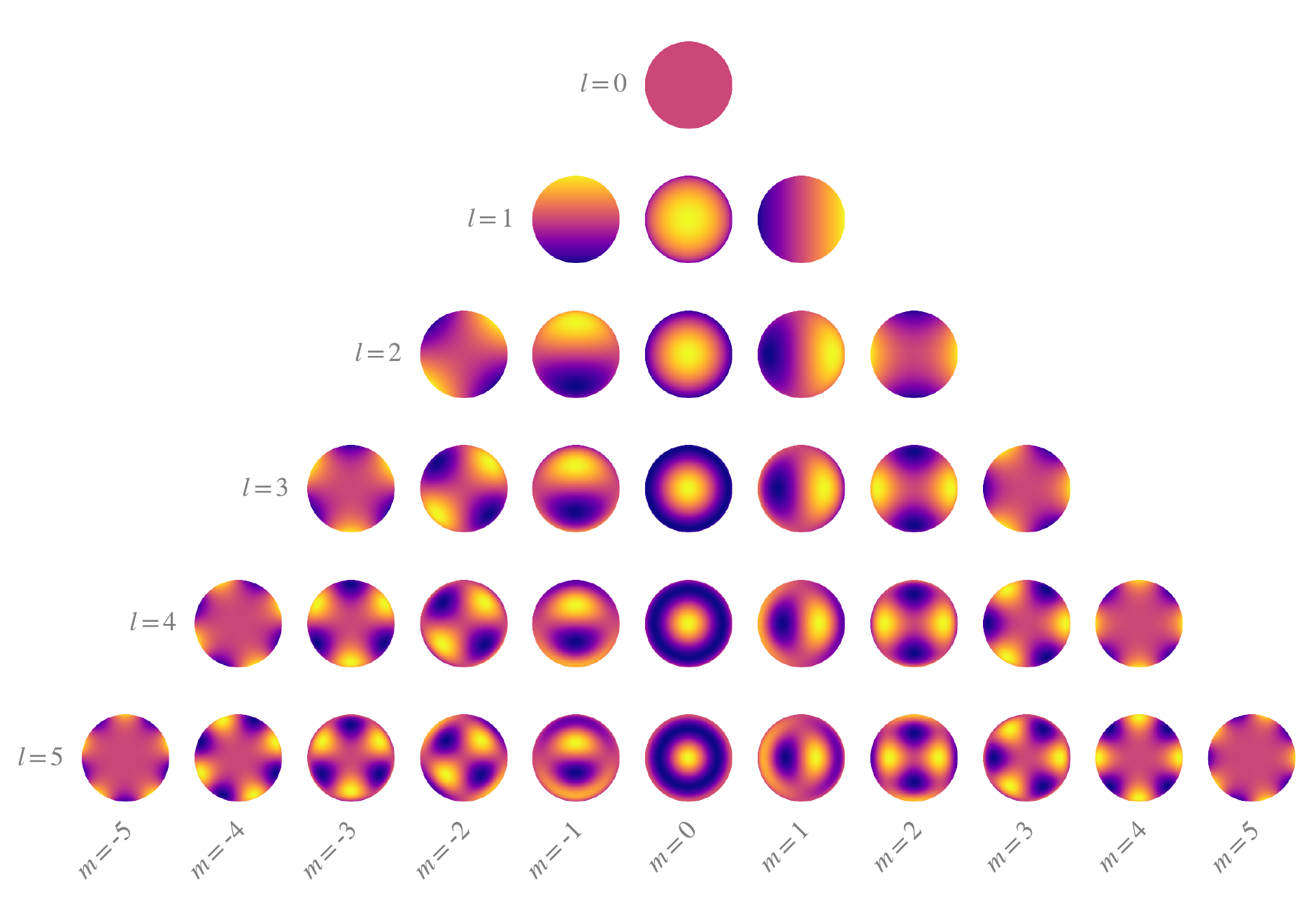}
        \oscaption{ylms}{%
            The real spherical harmonics in the polar frame
            ($\hat{\mathbf{x}}$ points to the right,
            $\hat{\mathbf{y}}$ points up, and $\hat{\mathbf{z}}$
            points out of the page)
            up to $l = 5$. Dark colors correspond to negative intensity
            and bright colors to positive intensity;
            rows correspond to the degree $l$ and columns to
            the order $m$. The set of all spherical harmonics forms a
            complete, orthogonal basis on the sphere.
            \label{fig:ylms}
        }
    \end{centering}
\end{figure}

While the stellar surfaces are all distinct, containing between one (top)
and six (bottom) large dark spots,
\textbf{their rotational light curves are identical}
(lower panel). This is true even in the absence of measurement
error: the mapping from a stellar surface
to its rotational light curve is so degenerate that there exist an infinite number of solutions
to the inverse problem. This fact has been pointed out recently in different contexts
\citep[e.g.,][]{Cowan2013,Rauscher2018,Sandford2019,Luger2019,Basri2020}, but it dates back at least to
\citet{Russell1906}, who demonstrated it by expanding the surface
intensity of a celestial body in terms of spherical harmonics
(see Figure~\ref{fig:ylms}). \citet{Russell1906} showed
that many of the modes comprising the intensity profile of a spherical
object are in the \emph{null space}, the set of surface features that have identically
zero effect on the light curve. In fact, as we will show in \S\ref{sec:nullspace},
the \emph{vast majority} of the modes are in the null space for rotational
light curves of stars. This is what allows us to construct pathological
scenarios like that shown in Figure~\ref{fig:degeneracies}, where the light curve
could be explained by any number of spots atop a heterogeneous bright background.

Stellar mapping studies tackle these degeneracies in different ways, but
it usually comes down to a choice of prior: when the data is not
sufficiently informative, assumptions---either implicit or explicit---are needed to discriminate
between competing solutions. In discrete spot models like the ones
discussed above, the degeneracy-breaking prior is (typically) the assumption that the
spots must be circular, have uniform contrast, and sit atop an otherwise
uniform photosphere. In gridded stellar surface models, it is common to
assume a regularization prior such as the maximum entropy penalty
\citep[e.g.,][]{Vogt1987}, which typically favors solutions with the fewest
number of dark pixels (usually referred to as the ``simplest'' solution).

While these assumptions may be approximately valid in some cases, it is
important to bear in mind that because of the light curve degeneracies discussed above,
\textbf{most of the information about the stellar surface usually comes from the modeling assumptions},
so it is very important to get these assumptions right. In general, starspots are not circular and do not have uniform
contrast throughout; nor do spots always arrange themselves in the highest
entropy configuration. The amount of bias introduced by these assumptions
will in general vary, but in principle it could be quite significant.


The goal of this paper is to explore the degeneracies at play in the
stellar surface mapping problem from a theoretical standpoint.
We will focus in particular on two sources of degeneracies: the
null space intrinsic to the mapping from a two-dimensional surface to
a one-dimensional light curve (\S\ref{sec:nullspace})
and the degeneracy due to the unknowability of the true normalization in
single-band photometry (\S\ref{sec:normalization}). Within each section we
will discuss ways to either break these degeneracies or marginalize
over unknowable quantities.
We will focus in particular on the power of ensemble analyses: the
joint analysis of many light curves of statistically ``similar''
stars. We will show that even though individual light curves are not very
constraining, light curves of many stars
observed at different inclinations can uniquely constrain certain
properties of the surfaces of those stars.
This idea was recently explored to some extent
in \citet{Morris2020}, who used ensemble analyses to derive constraints on
spot coverage areas as a function of stellar age. However, one of
the main conclusions of the present paper is that
quantities like \textbf{the
    total spot coverage and the total number of
    spots are not direct observables in single-band photometry.} Instead,
any constraints placed on these quantities, even in the context of
ensemble analyses, are usually driven by the choice of prior and other assumptions.

The present paper is also similar to \citet{Walkowicz2013} and
\citet{Basri2020}, who explored the information content of
stellar light curves from a large set of simulated spotted stellar
surfaces. While our paper is largely complementary to
that work, we instead approach the information content problem from
a theoretical---as opposed to empirical---point of view.
The present paper is the first in a series dedicated to the development of
techniques to perform robust inference about stellar surfaces from
unresolved photometric and spectroscopic measurements. The results
of this paper serve as the starting point for the development of
an interpretable Gaussian process for the ensemble analysis of
stellar light curves, which is the subject of
\citepalias[\citealt{PaperII}, hereafter][]{PaperII}.

\vspace{1em}

All of the figures in this paper were auto-generated
using the Azure Pipelines continuous integration (CI) service.
Icons next to each of the figures \codeicon \,
link to the exact script used to generate them to ensure the reproducibility
of our results. In this paper we also introduce the concept of equation
``unit tests'': \textsf{pytest}-compatible test scripts associated
with the principal equations that pass (fail) if the equation is correct (wrong),
in which case a clickable \testpassicon \, (\testfailicon) is shown next to the equation
label.
In most cases, the validity of an equation is gauged by comparison to
a numerical solution. Like the figure scripts, the equation unit tests are
run on Azure Pipelines upon every commit of the code.%
\footnote{
    These unit tests are certainly not foolproof: in particular, there is
    no guarantee against a mismatch in the \LaTeX \, version of an equation
    and its \Python implementation (e.g., due to an uncaught typo). However, they
    \emph{do} ensure that the linked \Python implementation is correct to
    within the accuracy of the numerical solution, providing readers with a
    valid implementation of the equation for purposes of reproducibility.
}

\section{The null space}
\label{sec:nullspace}

In this section, we define the concept of the \emph{null space}
and present some demonstrations showing how it can affect inferences
about stellar surfaces.
For simplicity, we assume we know quantities like the stellar
inclination, rotation period, and limb darkening coefficients
exactly, and we assume the stellar surface does not vary in time
(i.e., spots do not evolve and the star rotates as a rigid body).
Because of these assumptions, our constraints on the information
content of light curves should be viewed as strict \emph{upper limits}.

We also present some examples of how the degeneracies due to
the null space can be tackled.
While our examples below may appear somewhat idealized, at the end
of this section we discuss how our ideas generalize to more
realistic scenario; we also revisit these assumptions in \S\ref{sec:caveats}.
This is also a topic that we discuss in much more detail in \citetalias{PaperII}.

\subsection{Rank of the flux operator}
In general, inferring all of the properties of a stellar surface from its light curve alone
is not
only difficult, but \emph{formally impossible}. To understand why, consider
an expansion of the stellar surface intensity in the spherical
harmonic basis out to arbitrary order%
\footnote{%
    We note that this expansion is fully general, since spherical harmonics
    constitute a complete basis on the sphere.
} (see Figure~\ref{fig:ylms}).
Assuming (for the moment) that the
star rotates about an axis that points up along the page, the observed
light curve may be expressed as a weighted sum of the disk-integrated intensity
of each of the spherical harmonics as they rotate about that same axis \citep{Luger2019}.
However, not all spherical harmonics will contribute to the full light curve,
as many (in fact, most) of the spherical harmonics are perfectly antisymmetric
about the equator. This is the case for
the $l = 1$, $m = -1$ harmonic, which integrates to zero regardless of
the phase at which it is viewed. The same is true, in fact, for all other harmonics
of order $m = -1$ and (perhaps less obviously) for all harmonics with odd
$l = 3, 5, 7, ...$ Furthermore, there exist many linear combinations of
spherical harmonics that similarly integrate to zero at all rotational
phases. Together, these modes constitute the \emph{null space} of the problem:
the set of modes on the surface that do not contribute to the observed
light curve and therefore cannot be probed from photometry.

For rotational light curves,
the vast majority of the surface modes lie in the null
space. To show this, we will make use of the fact that we can express
the vector of $K$ observed fluxes $\mathbf{f}$ (i.e., the light curve)
as a linear operation on the vector of $N$
spherical harmonic coefficients $\mathbf{y}$ \citep{Luger2019}:
\begin{align}
    \label{eq:fAy}
    \mathbf{f} = \mathbf{1} + \pmb{\mathcal{A}} \, \mathbf{y}
    \quad,
\end{align}
where $\pmb{\mathcal{A}}$ is the $(K \times N)$ \emph{design matrix} of the transformation, whose columns
describe how each of the $N$ components in the spherical harmonic basis contribute
to each of the $K$ points in the light curve.%
\footnote{%
    For rotational light curves, the rows of $\pmb{\mathcal{A}}$ are given by the
    quantity $\mathbf{r}^\top \pmb{\mathcal{A}}_1 \mathbf{R}$ in Equation~(18) of
    \citet{Luger2019}, where $\mathbf{r}^\top$ is a vector of disk-integrated
    intensities, $\pmb{\mathcal{A}}_1$ is a change of basis matrix, and $\mathbf{R}$
    is a spherical harmonic rotation matrix that depends on the stellar inclination
    and the rotational phase of the star. In Equation~(\ref{eq:fAy}) we explicitly
    add a vector of ones to enforce a unit baseline for spotless stars.
    Refer to \citet{Luger2019} for more details.
}
Even though we are explicitly choosing the spherical harmonics as the basis in which
we describe the stellar surface, Equation~(\ref{eq:fAy}) is quite general and applies to
\emph{any} basis that is linearly related to the flux. For instance, $\mathbf{y}$
could instead describe the intensity deficits in each of the $N$ pixels of a gridded stellar
surface, in which case $\pmb{\mathcal{A}}$ would be the matrix of pixel visibilities that describe
how to sum each of the pixels to obtain the observed light curve.

The size of the null space is called the \emph{nullity}, and it is equal to
$N - R$, where $N$ is once again the number of coefficients describing the
stellar surface and $R$ is the \emph{rank} of the flux operator $\pmb{\mathcal{A}}$.
The rank $R$ is the number of linearly independent columns in $\pmb{\mathcal{A}}$, which
can be computed numerically using any standard linear algebra package. It is equal
to the number of independent components that can be measured given
an observation of $\mathbf{f}$.

\begin{figure}[t!]
    \begin{centering}
        \includegraphics[width=\linewidth]{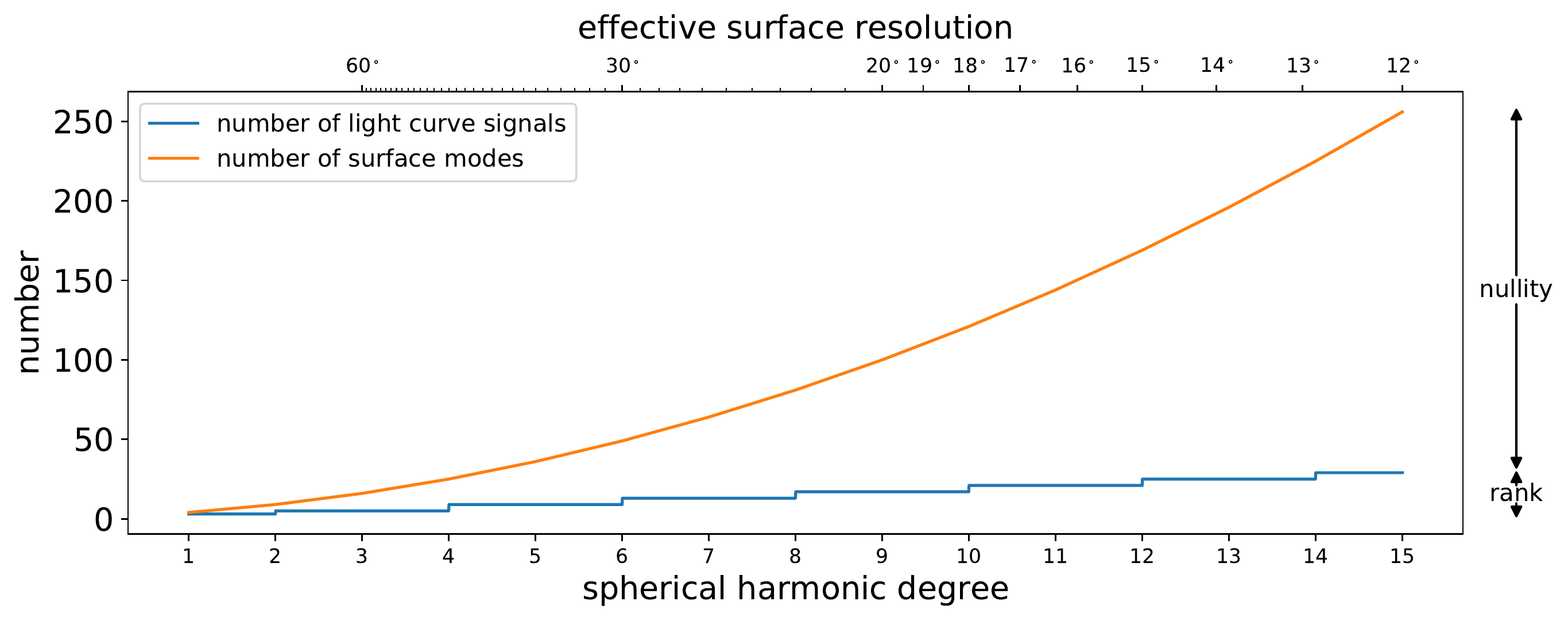}
        \oscaption{rank}{%
            Rank and nullity of the flux operator. The orange curve
            shows the number of spherical harmonic coefficients required
            to fully describe a stellar surface up to a given spherical
            harmonic degree (bottom axis) or, equivalently, up to an
            effective surface resolution (top axis). The blue curve shows
            the rank of the flux operator, corresponding to the maximum
            number of independent degrees of freedom that can be constrained from a
            light curve. The size of the null space (the nullity) is
            the difference between the two curves.
            \label{fig:rank}
        }
    \end{centering}
\end{figure}

Figure~\ref{fig:rank} shows the rank and nullity of the flux operator
as a function of the resolution of the surface map (quantified as the
spherical harmonic degree $l$ of the expansion). The orange
curve is the number of spherical harmonic coefficients needed to
represent a surface map up to degree $l$, and is equal to $N = (l + 1)^2$.
The blue line shows the rank of the flux operator $\pmb{\mathcal{A}}$ as a function
of $l$, which scales as $R \approx 2l + 1$. The nullity is
simply the difference between $N$ and $R$.

The most striking feature in Figure~\ref{fig:rank} is how quickly the two
curves diverge as $l$ increases. What this means for the mapping problem
is that the number of surface modes---the total information needed to
represent a surface map at some resolution---grows much more quickly than the
number of independent degrees of freedom in the light curve.
At all but the lowest resolutions, there are always more features in the
null space than components one can measure in the light curve, a difference
that grows \emph{quadratically} with $l$.
This means that although a light curve can tell us some information about
a stellar surface on very large scales, the amount of information it tells
us quickly decreases for smaller scales and all but vanishes for the smallest
surface features.

It is also worth noting the piecewise nature of the rank as a function of
$l$: increasing the degree of the expansion from even $l$ to odd $l$ does not
increase the rank of the flux operator. Put another way, odd spherical
harmonic modes with $l > 1$ are always in the null space of the light curve.
We will return to this point below.

\subsection{Decomposition of the flux operator}
\label{sec:svd}
In Appendix~\ref{sec:app-svd} we show that it is straightforward to
construct linear operators $\mathbf{P}$ and $\mathbf{N}$ from the
design matrix $\pmb{\mathcal{A}}$ such that
\begin{align}
    \mathbf{y}_\bullet & = \mathbf{P} \, \mathbf{y}
\end{align}
is the component of the surface map in the \emph{preimage} and
\begin{align}
    \mathbf{y}_\circ & = \mathbf{N} \, \mathbf{y}
\end{align}
is the component of the map in the \emph{null space}.
The \emph{preimage operator} $\mathbf{P}$
transforms a vector $\mathbf{y}$ in the surface map basis in such a way
that it preserves the information in $\mathbf{y}$ that gets mapped
onto the light curve $\mathbf{f}$ via $\pmb{\mathcal{A}}$ (the \emph{preimage}) and discards the rest. The
\emph{null space operator} $\mathbf{N}$ does the opposite: it preserves only
the information in $\mathbf{y}$ that gets mapped onto the zero
vector via $\pmb{\mathcal{A}}$ (the \emph{null space}).
In other words,
the $\mathbf{P}$ and $\mathbf{N}$ operators reveal the
components of the surface map that contribute to the
light curve ($\mathbf{y}_\bullet$) and the components that don't ($\mathbf{y}_\circ$).
The vector $\mathbf{y}_\bullet$ represents all the information that can be
learned from a stellar light curve, while $\mathbf{y}_\circ$ represents all the
information that cannot.

\begin{figure}[p!]
    \begin{centering}
        \vspace{-4em}
        \includegraphics[width=1in]{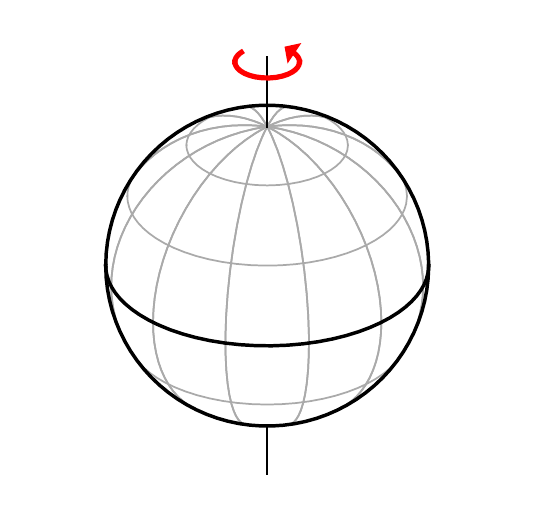}
        \\[0.5em]
        \includegraphics[width=\linewidth]{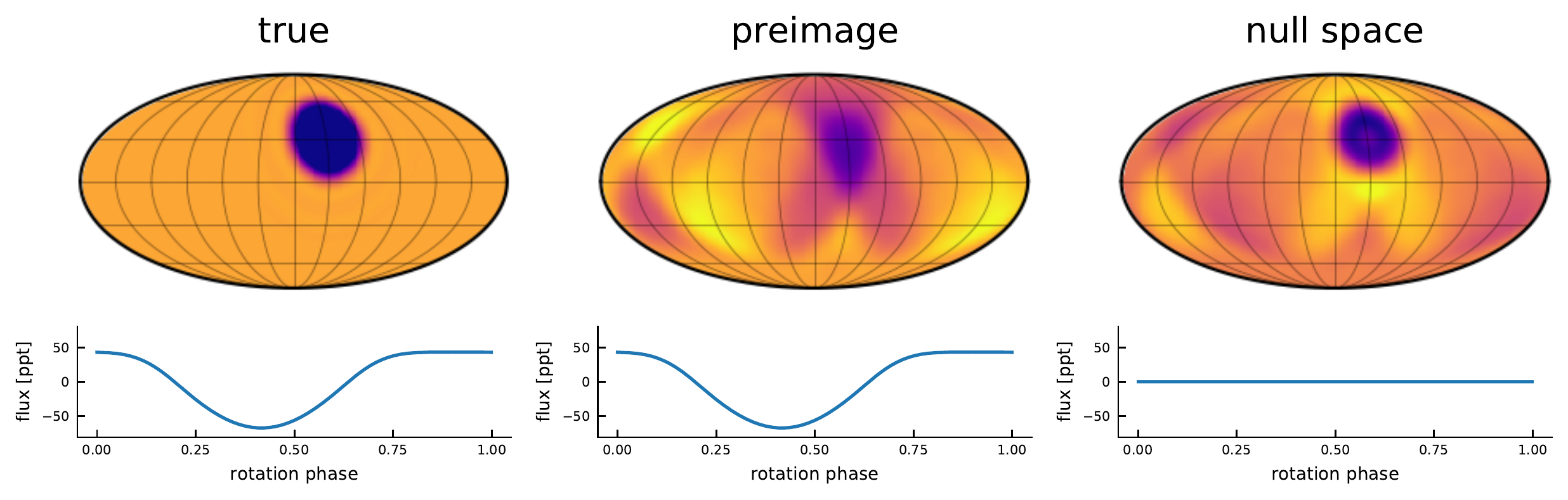}
        \\[1em]
        \includegraphics[width=\linewidth]{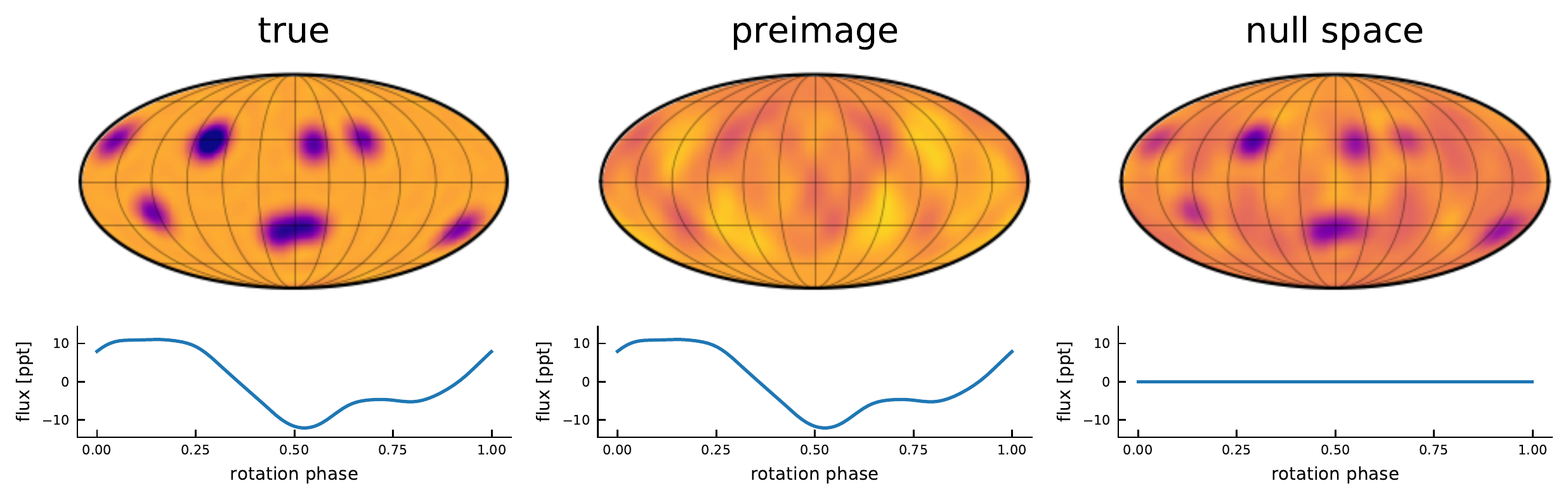}
        \\[1em]
        \includegraphics[width=\linewidth]{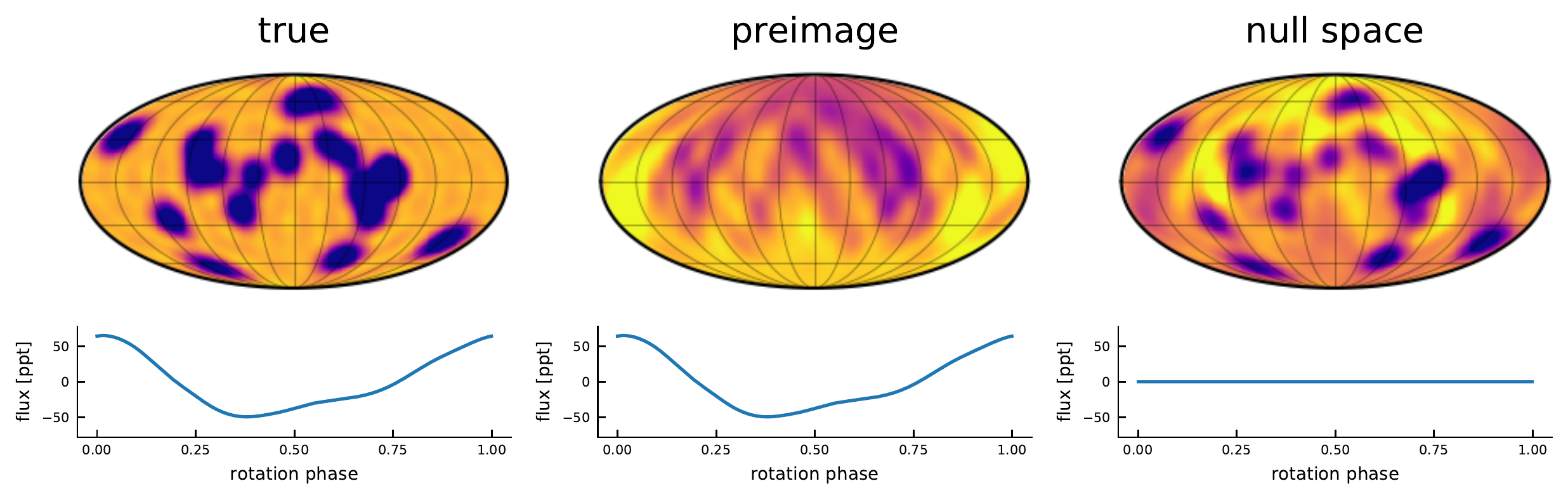}
        \\[1em]
        \includegraphics[width=\linewidth]{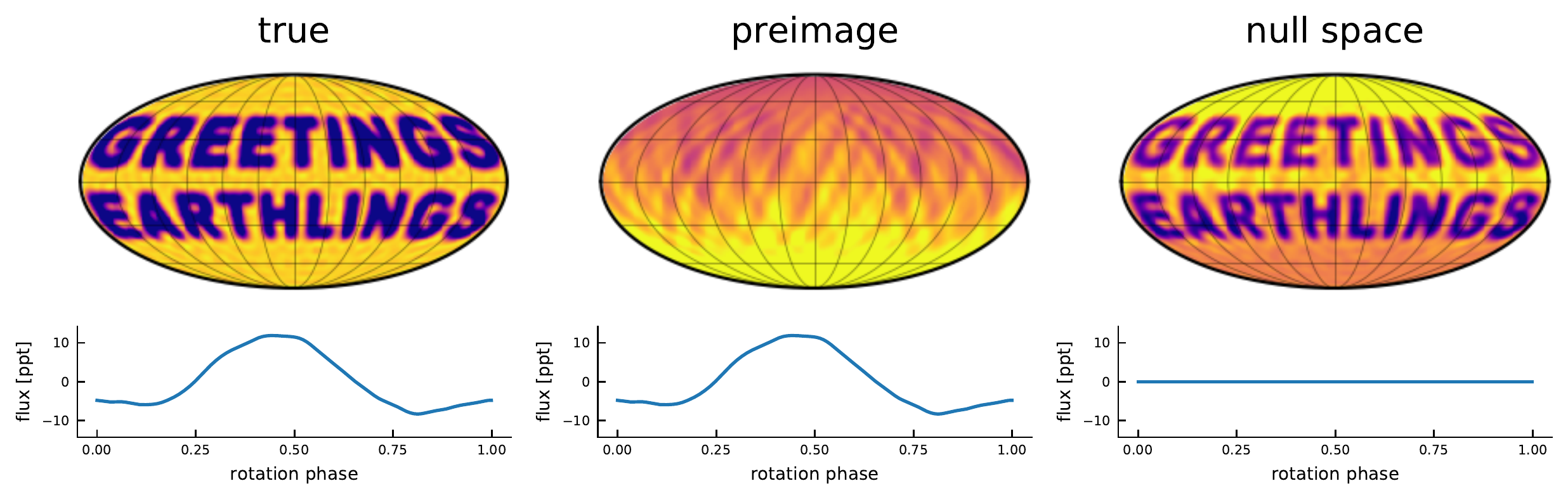}
        \oscaption{nullspace_preimage}{%
            Decomposition of a surface map (left column) into its
            preimage (center) and null space (right) components
            for different surfaces, and their corresponding contributions
            to the rotational light curve. The preimage is the
            set of surfaces modes that map onto the light curve;
            the null space is the set of modes that do not.
            An inclination of $60^\circ$ is assumed when computing the flux.
            The vast majority of surface modes are in the null space
            of the light curve problem and therefore do not contribute
            to the observed flux.
            \label{fig:nullspace_preimage}
        }
    \end{centering}
\end{figure}

\begin{figure}[p!]
    \begin{centering}
        \vspace{-4em}
        \includegraphics[width=1in]{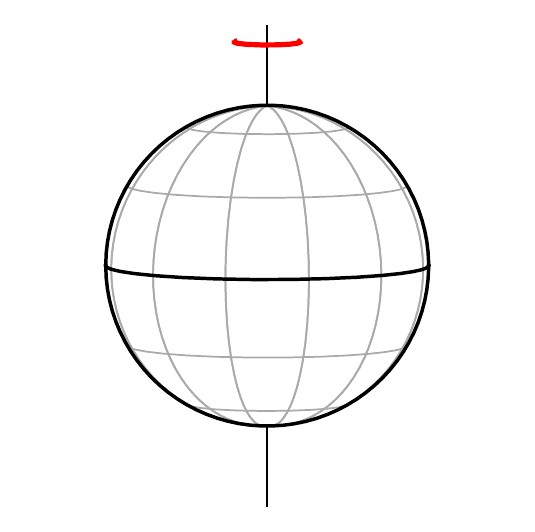}
        \\[0.5em]
        \includegraphics[width=\linewidth]{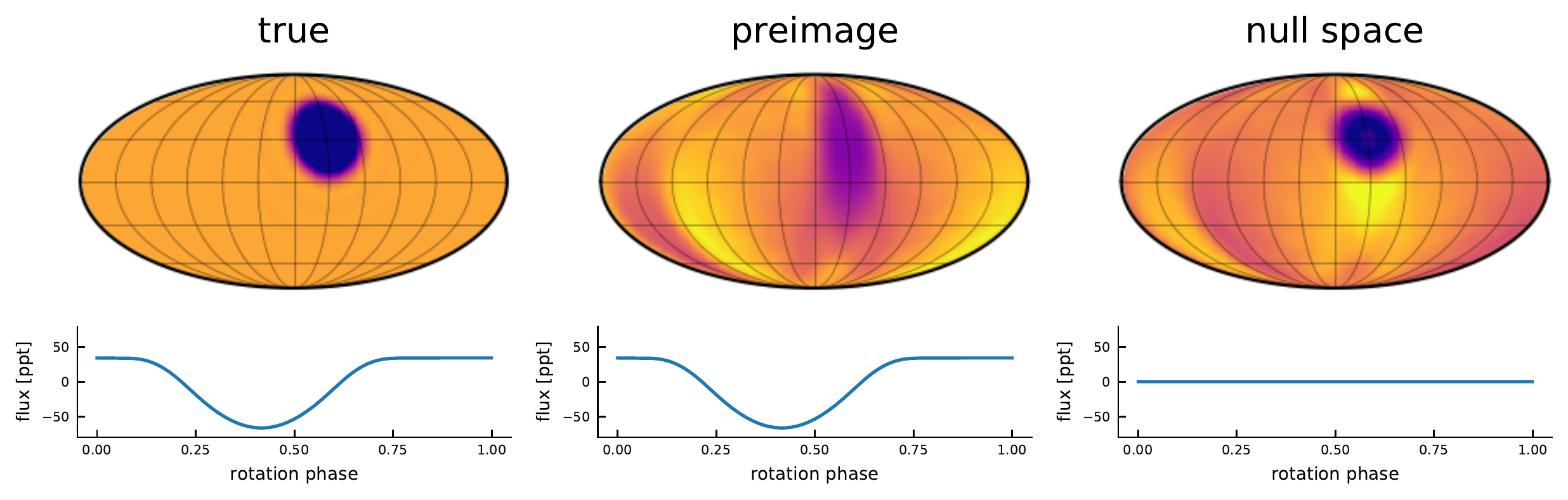}
        \\[1em]
        \includegraphics[width=\linewidth]{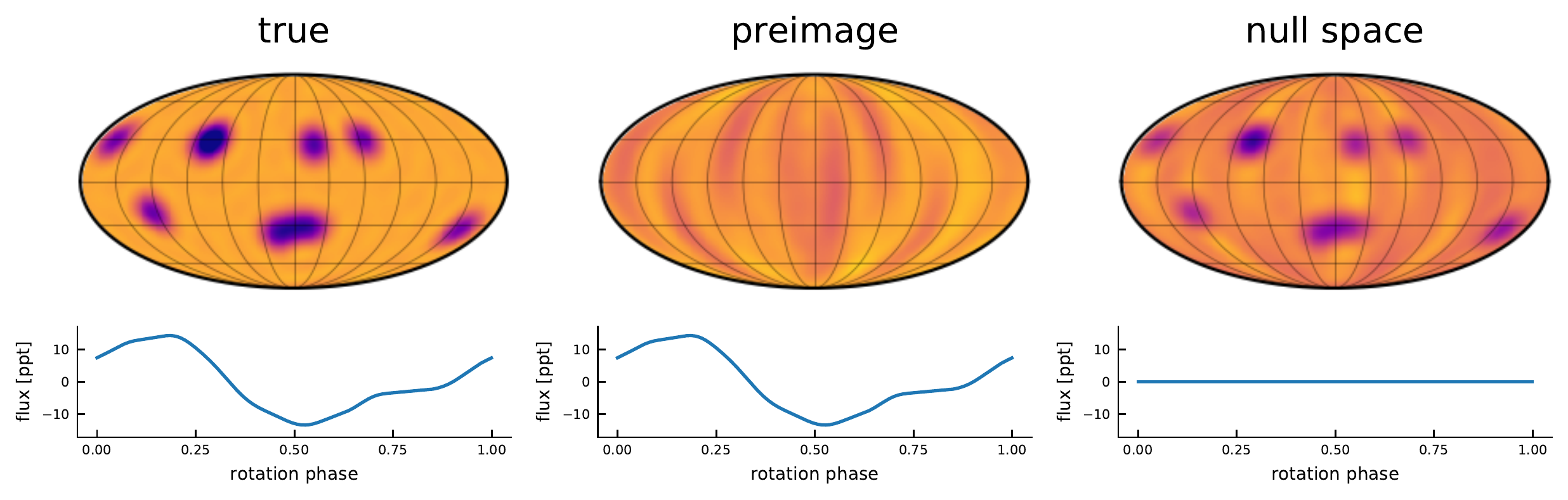}
        \\[1em]
        \includegraphics[width=\linewidth]{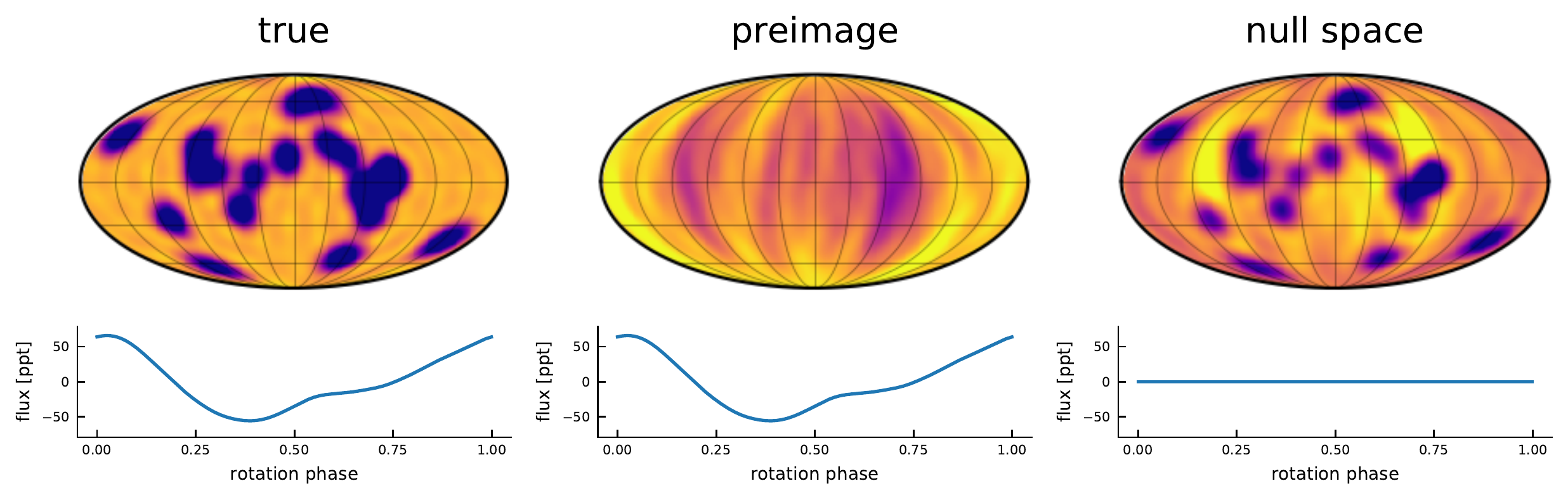}
        \\[1em]
        \includegraphics[width=\linewidth]{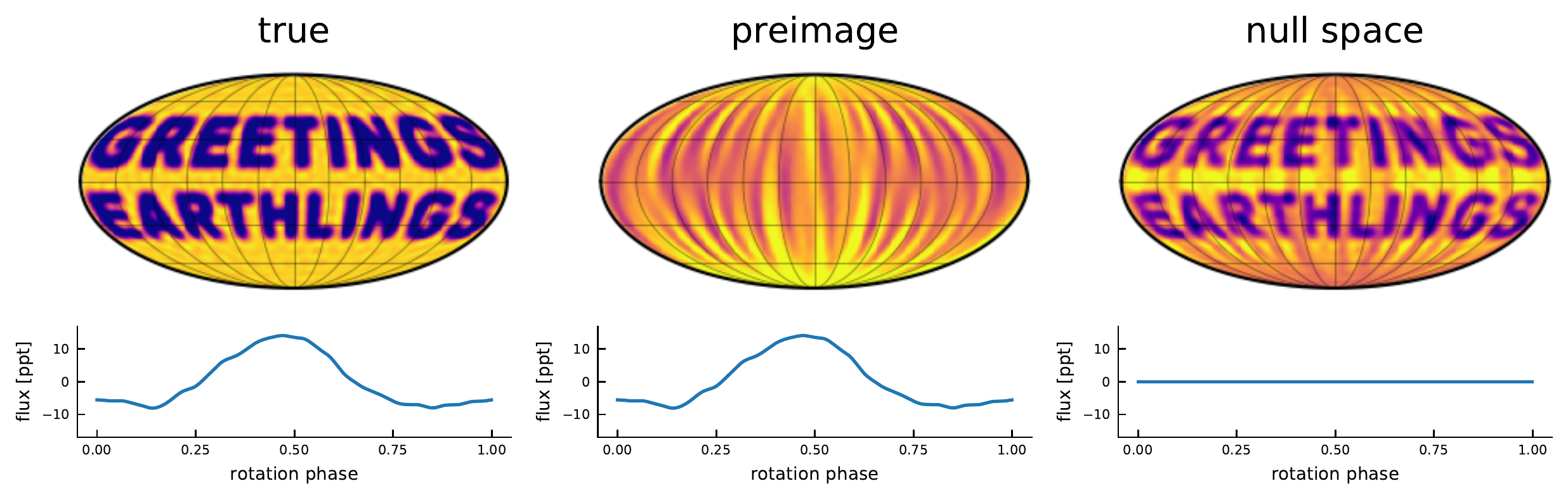}
        \oscaption{nullspace_preimage}{%
            Same as Figure~\ref{fig:nullspace_preimage}, but for
            a stellar inclination of $85^\circ$. As the stellar rotation vector
            becomes perpendicular to the line of sight, it becomes
            more difficult to constrain latitudinal information.
            \label{fig:nullspace_preimage_85}
        }
    \end{centering}
\end{figure}

It is instructive to visualize these components in
an actual surface mapping exercise. Figure~\ref{fig:nullspace_preimage}
shows the decomposition of four hypothetical surfaces (left
column) into preimage (center) and null space (right)
components under the flux operator $\pmb{\mathcal{A}}$, which we compute for
definiteness at an inclination of $I = 60^\circ$ over a full rotation.
Surfaces are shown in an equal-area Mollweide projection alongside
the corresponding light curves.
Note that both the true map and its associated light curve are simply equal to
the sum of the preimage and null space components (Appendix~\ref{sec:app-svd}).
As expected, all of the information in the light curve comes
from the preimage, and the null space contributes exactly zero
flux at all phases. However, most of the information about the
\emph{surface} is stuck in the null space!

In the top row of the figure, corresponding to a surface with
a single large spot, it is clear that the light curve
contains information about the presence of the spot at roughly the
correct longitude and latitude. There are additional artefacts
across the stellar surface, and the exact shape of the spot is
not well constrained by the data; these issues, however, can
easily be resolved with a circular spot prior.

The degeneracies of the mapping problem are much more apparent
in the second and third rows, corresponding to surfaces with
many, smaller spots. The locations, sizes, and the very existence
of most of the spots are simply not encoded in the light curve.
Even with an extremely restrictive prior, it may be difficult---if
not impossible---to learn the properties of the spots
from the individual light curves.

The final row corresponds to a surface with much higher resolution
features; this example highlights how the information content of
light curves all but vanishes at small scales. Virtually all of the
spatial information at the scales of interest is in the null space.

\subsection{Implications for inference}
The model mapping spherical harmonic coefficients to observed fluxes described
in the preceding sections is linear, so we might expect that we could fit
for the spherical harmonics using linear least-squares
\begin{align}
    \label{eq:linlstsq}
    \hat{\mathbf{y}} & \stackrel{?}{=}
    \left(\pmb{\mathcal{A}}^\top\,\pmb{\mathcal{A}}\right)^{-1}\,\pmb{\mathcal{A}}^\top\,\left(\mathbf{f} - \mathbf{1}\right)
    \quad.
\end{align}
But, since the flux operator $\pmb{\mathcal{A}}$ is low-rank, the model is
underdetermined, so the above operation is not defined and, equivalently,
the Fisher information matrix
\begin{align}
    \label{eq:fisher-info}
    \pmb{\mathcal{I}}(\mathbf{y}) & = \frac{1}{\sigma_f^2} \, \pmb{\mathcal{A}}^\top\,\pmb{\mathcal{A}}
\end{align}
is singular.
In Equation~(\ref{eq:fisher-info}), $\sigma_f$ is the flux measurement uncertainty.
There exist many methods and procedures for linear regression for underdetermined
models, but all solutions amount to imposing stronger assumptions about the
model in order to break the degeneracies.

One standard method is regularized least squares, in which
Equation~(\ref{eq:linlstsq}) becomes
\begin{align}
    \hat{\mathbf{y}}_\lambda & \equiv
    \left(\pmb{\mathcal{A}}^\top\,\pmb{\mathcal{A}} + \lambda\,\mathbf{I}\right)^{-1}\,\pmb{\mathcal{A}}^\top\,\left(\mathbf{f} - \mathbf{1}\right)
\end{align}
where $\mathbf{I}$ is the identity matrix and $\lambda$ is a parameter
controlling the strength of the regularization.
In a Bayesian context, this can be interpreted as placing a Gaussian prior
with variance $\sigma_0^{2} = \nicefrac{\sigma_f^2}{\lambda}$ on the spherical harmonic coefficients.

In the limit of infinitesimal regularization, it can be demonstrated
\citep[see][for example]{Hogg2021} that
\begin{proof}{test_svd_lstsq}
    \label{eq:inference}
    \lim_{\lambda \rightarrow 0_+} \hat{\mathbf{y}}_\lambda
    &= \mathbf{y}_\bullet
    \quad.
\end{proof}
In this sense, $\mathbf{y}_\bullet$ represents our knowledge about the
surface of a star after an observation if we have no prior information
whatsoever on $\mathbf{y}$.
\emph{Any other information about the surface is entirely driven by our assumptions.}%

Equation~(\ref{eq:inference}) is the solution to the surface map
we would obtain if we knew \emph{nothing} about the stellar surface
before analyzing the light curve. In practice, this is never really
the case. For instance, we know that stellar surfaces
must have non-negative intensities everywhere. While this may seem
like a trivial prior, non-negativity can be a powerful degeneracy-breaking
constraint \citep[e.g.,][]{Fienup1982}. Moreover, we know that
stellar surfaces usually consist of localized features; under an
appropriate compactness prior, solutions
like the preimage in the bottom row of
Figure~\ref{fig:nullspace_preimage} (for example) could be
confidently ruled out.

The effect of different modeling assumptions on the structure and size of the null
space is beyond the scope of this paper. In general, this is a particularly
difficult question to address because non-Gaussian priors on the
surface map break the linearity of the problem. While compactness constraints
(like the assumption of a small number of discrete circular spots) can
break many of the degeneracies discussed here, simulation-based
arguments show that it is still not possible to uniquely constrain
their number, locations, or sizes from individual light curves
\citep{Basri2020}.


\subsection{Dependence on inclination}
\label{sec:inclination}

Figure~\ref{fig:nullspace_preimage_85} shows the same decomposition
of stellar surfaces into what can be learned (the preimage) and what
cannot be learned (the null space) from a light cuve, but this time
for stars viewed at an inclination $I = 85^\circ$. Interestingly,
the structure of the null space is somewhat different; in particular,
features in the preimage are latitudinally smeared. This
issue is well known \citep[e.g.,][]{Cowan2009,Basri2020}, and any
information about which hemisphere a feature is in formally vanishes
as $I \rightarrow 90^\circ$.

Instead of lamenting the difficulties of constraining stellar surface features
at near-edge-on orientations, let us focus on the fact that the
null space is a function of inclination.
A useful property of
the spherical harmonics is that a rotation operation on any
component in the basis can only change the order $m$ of the
harmonic; the degree $l$ is constant under rotation. In other
words, rotating any of the spherical harmonics in Figure~\ref{fig:ylms}
about an arbitrary axis simply yields a weighted sum of the spherical
harmonics along its row.%
\footnote{For instance, rotation of $Y_{1,-1}$ by $90^\circ$ about
    $\hat{\mathbf{x}}$ yields $Y_{1,1}$; other rotations will in general
    yield a weighted combination of $Y_{1,-1}$, $Y_{1,0}$, and $Y_{1,1}$.}
This means that changing the inclination of the star---which changes
the axis of rotation in the observer's frame---simply changes the weighting
of modes that give rise to certain signals in the light curve.
This, in turn, results in the dependence of the null space on inclination.

To better understand this effect, let us define the quantity
\begin{align}
    S \equiv
    1 -
    \lim_{\lambda \rightarrow 0_+}
    \mathrm{diag}
    \left(
    \lambda
    \left(\pmb{\mathcal{A}}^\top\,\pmb{\mathcal{A}} + \lambda\,\mathbf{I}\right)^{-1}
    \right)
    \quad,
\end{align}
which we will refer to as the \emph{\shrinkage}
of the coefficients $\mathbf{y}$ characterizing the stellar surface.
Note that in a Bayesian context, this is
equivalent to the \emph{posterior shrinkage}
\begin{align}
    \label{eq:shrinkage}
    S = 1 - \lim\limits_{\sigma_0^2 \rightarrow \infty}
    \frac{\sigma^2}{\sigma_0^2}
    \quad,
\end{align}
where $\sigma_0^2$ is the prior variance
and $\sigma^2$ is the posterior variance
on a particular surface mode we are trying to constrain
\citep[see, e.g.,][]{Betancourt2018}.

The \shrinkage is a dimensionless quantity describing how informative a measurement is about a
given mode on the surface in the limit of infinite signal-to-noise ratio (SNR)
and is independent of what the stellar surface actually looks like.
If, at infinite SNR and with no regularization
(or, in a Bayesian context, a completely uninformative prior),
a particular mode can be learned exactly from a dataset, the \shrinkage
is defined to be unity. Conversely, if the data is completely
unconstraining of that mode (i.e., it is entirely in the null space),
$S$ will tend to zero.

\begin{figure}[t!]
    \begin{centering}
        \includegraphics[width=\linewidth]{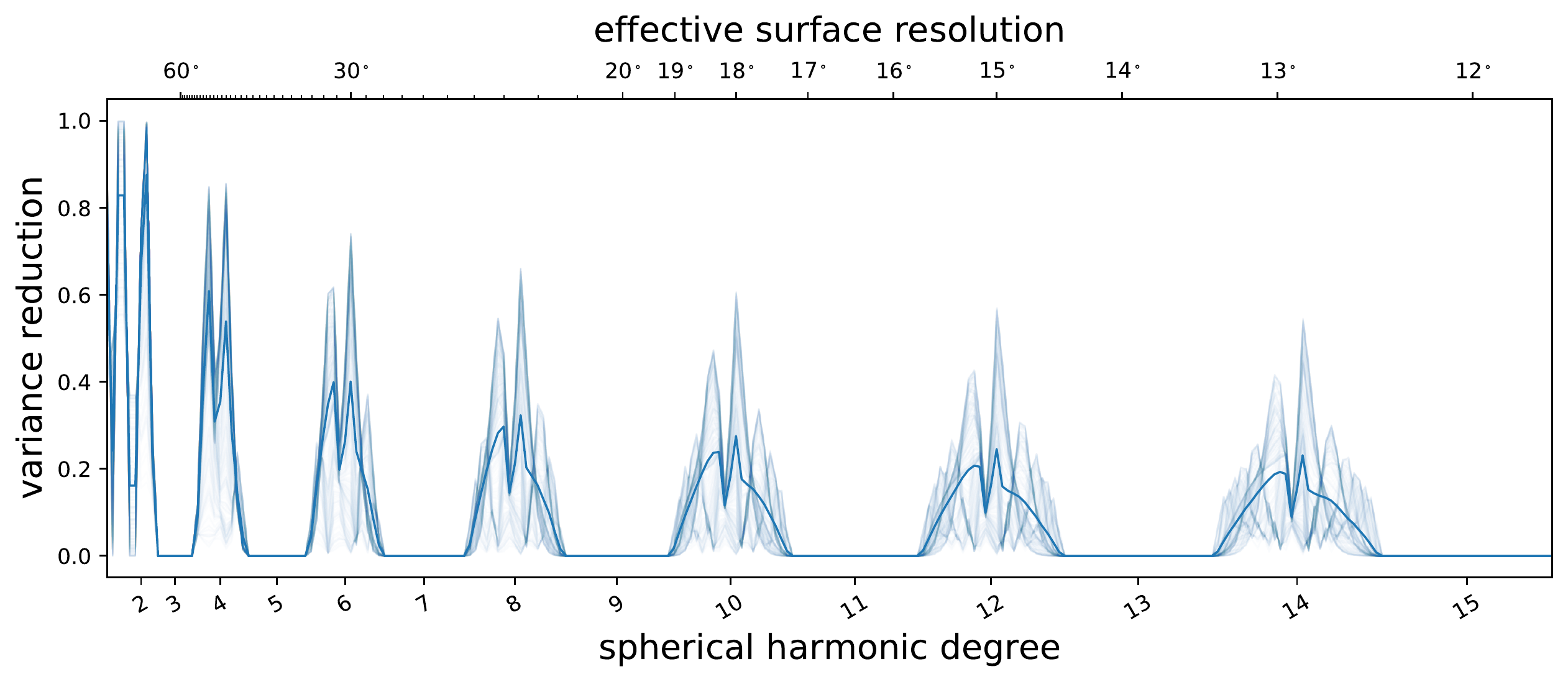}
        \oscaption{nullspace_ensemble_single}{%
            The \shrinkage as a function of spherical harmonic
            degree $l$ given a single observation of a star at a random
            inclination (thin blue curves). The mean \shrinkage is shown
            as the thicker curve. The information content of light curves
            tends to zero as $l$ increases, and odd $l > 1$ modes are in the
            null space at all inclinations.
            \label{fig:nullspace_ensemble_single}
        }
    \end{centering}
\end{figure}

Figure~\ref{fig:nullspace_ensemble_single} shows the \shrinkage $S$
given a single observation of a star at a random inclination.
Each thin blue curve corresponds to a particular draw from an isotropic
inclination distribution; the thick blue curve is the average over 300 trials.
For some of the low-degree modes, $S$ is relatively high: it is
fairly easy to constrain the dipole moment from a light curve, as this is
usually the dominant sinusoidal signal. However, as the degree $l$
increases, $S$ decreases dramatically: at $l = 14$, corresponding
to features on scales of roughly $13^\circ$, the light curve
can only tell us about $\sim 10\%$ of the total information about what the
surface looks like. As $l$ increases further, $S$ tends to zero.
Another important feature of $S$, which we hinted at above,
is that it is exactly zero for
all odd-degree modes above $l = 1$. This is a well-known fact: all odd spherical
harmonics other than the dipole are in the null space \emph{regardless of
    inclination} \citep[e.g.,][]{Luger2019}. In other words, these spherical
harmonics are perfectly antisymmetric in projection over the unit disk
when viewed from any orientation. Absent structure to break these symmetries
(see below), we simply cannot learn anything about these modes from
stellar light curves. If we average over $S$ for all modes up to $l=15$,
we find that a single light curve measurement can only tell us $\sim 9\%$
of the information about the surface on those scales.

Fortunately, however, there is quite a bit of scatter in $S$
for different values of the stellar inclination.
As we will see, we can use this dependence of the null space on inclination to our
advantage. If we could observe a star from many different vantage points,
we would be able to break many of the degeneracies at play, since we
would get different constraints on the amplitude of each mode when viewed
at different inclinations. This, of course, is not possible (at least not
yet!). But what we \emph{can} do is observe many similar stars, each viewed
at a different (random) inclination, and attempt to learn something about
the properties of the ensemble of stars as a whole.
In the
following section, we explore the role of ensemble analyses in breaking
the degeneracies of the mapping problem in more detail.

\subsection{Ensemble analyses}
\label{sec:ensemble}

\begin{figure}[t!]
    \begin{centering}
        \includegraphics[width=\linewidth]{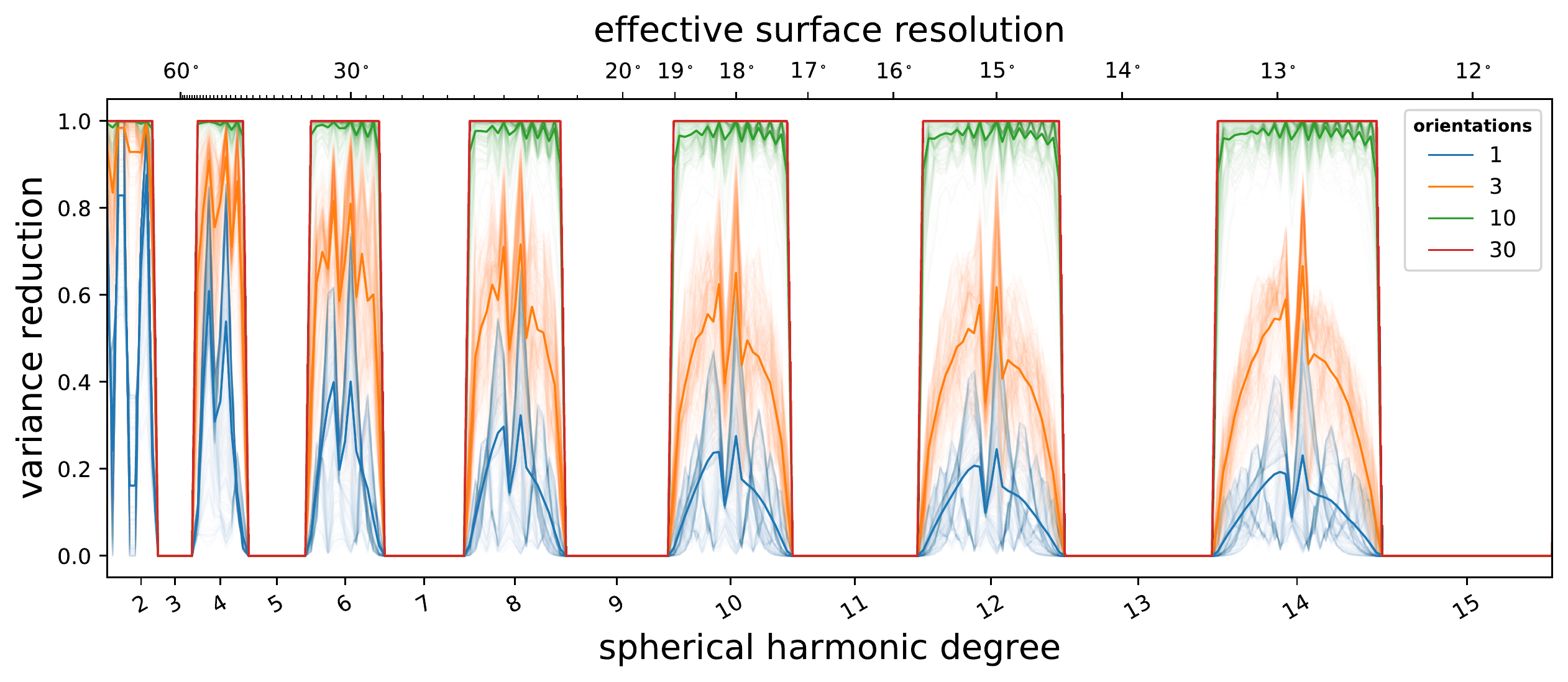}
        \oscaption{nullspace_ensemble}{%
            Similar to Figure~\ref{fig:nullspace_ensemble_single}, but
            assuming the observer can measure the light curves
            of 1 (blue), 3 (orange), 10 (green), and 30 (red) \emph{identical}
            stars--(same surface map, same rotational phase--but viewed at random orientations.
            As we saw in Figure~\ref{fig:nullspace_ensemble_single},
            the information content in the light curve of a star observed
            from a single vantage point approaches zero as $l$
            increases. However, observing many identical stars from different vantage points
            allows one to recover nearly all of the information in the
            even spherical harmonic modes. This is why ensemble analyses of
            many similar stars at different inclinations allows us to infer
            their surface properties.
            \label{fig:nullspace_ensemble}
        }
    \end{centering}
\end{figure}

In an ensemble analysis, we assume we observe the lightcurves of many stars
that are ``similar'' in some statistical sense. As a thought experiment,
let us consider an extreme version of ensemble analysis in which all the
stars in our sample happen to have \emph{identical} surfaces. We will
still assume they are oriented at random inclinations, as we would expect for
field stars.
Figure~\ref{fig:nullspace_ensemble} shows the \shrinkage
curves for this hypothetical scenario, assuming we have access to
light curves of 1 (blue), 3 (orange), 10 (green) and 30 (red)
identical stars viewed at random inclinations.

The addition of light curve measurements at different orientations
increases the \shrinkage at all even spherical harmonic degrees
(the odd degrees, as we mentioned above, are always invisible).
Note that since we are in the limit of infinite SNR, the fact that we have
more light curves (i.e., more data) is irrelevant: the
increase in the \shrinkage is instead due to the fact that our observations
from different vantage points broke some degeneracies in the problem.
This is a consequence of the fact we mentioned in the previous section:
the null space (for the even modes)
is a strong function of the inclination.

If we average over all modes, we obtain an average $S$ of $\sim 24\%$
for $l\leq15$ when our sample size is 3 (orange curves): we have more than doubled the
information content of our observations. If we further increase our sample size
to 10 (green curves), the \shrinkage approaches $100\%$ for all even $l\leq15$
modes, effectively saturating for a sample size of 30 (red curves).
Thus, if we were able to measure light curves of identical stars
from many different inclinations, the null space would consist \emph{only}
of the odd modes. In the limit of a large number of light curves,
and assuming all stars in the sample have identical surfaces,
\textbf{ensemble analyses can tell us up to
    half of all the information about the stellar surfaces.}

\begin{figure}[p!]
    \begin{centering}
        \vspace{-4em}
        \includegraphics[width=1in]{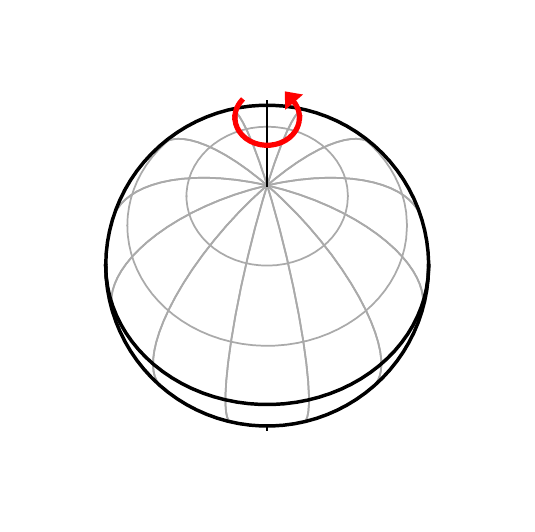}
        \includegraphics[width=1in]{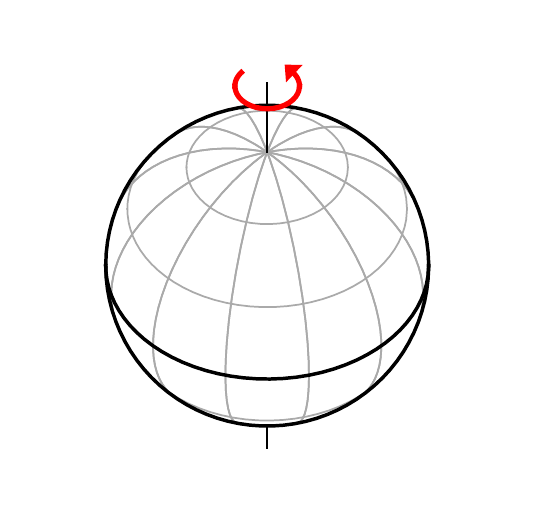}
        \includegraphics[width=1in]{figures/wireframe_60.pdf}
        \includegraphics[width=1in]{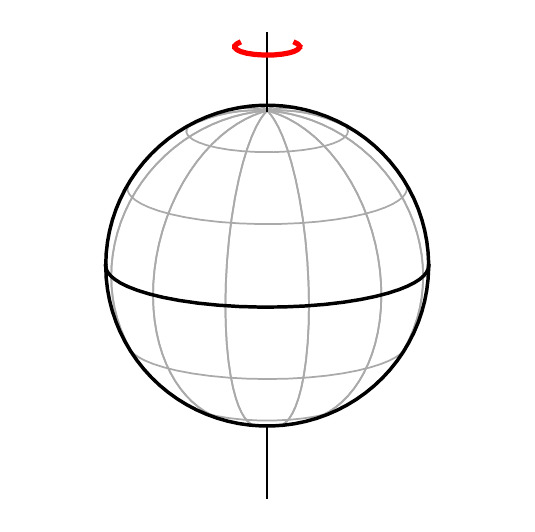}
        \\[0.5em]
        \includegraphics[width=\linewidth]{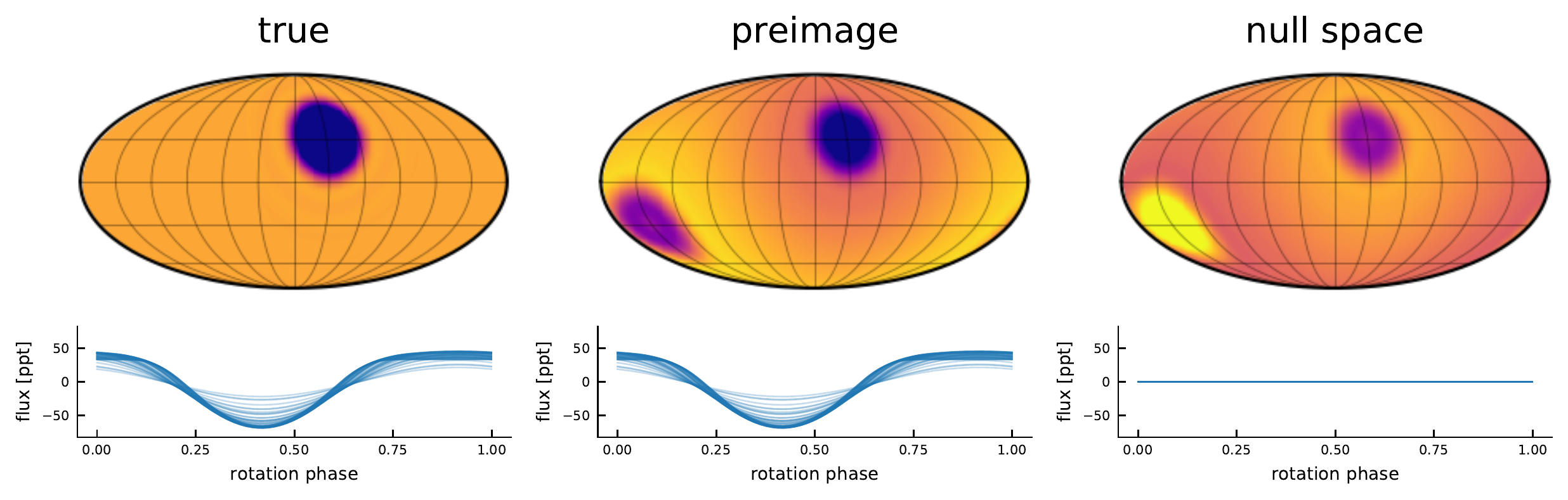}
        \\[1em]
        \includegraphics[width=\linewidth]{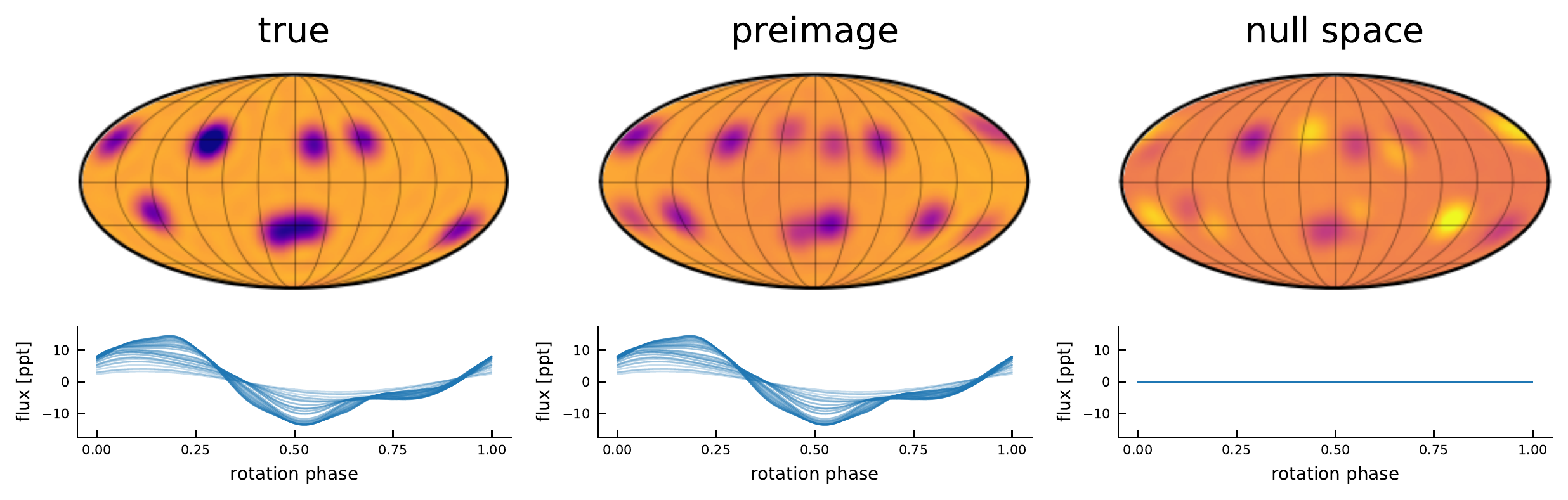}
        \\[1em]
        \includegraphics[width=\linewidth]{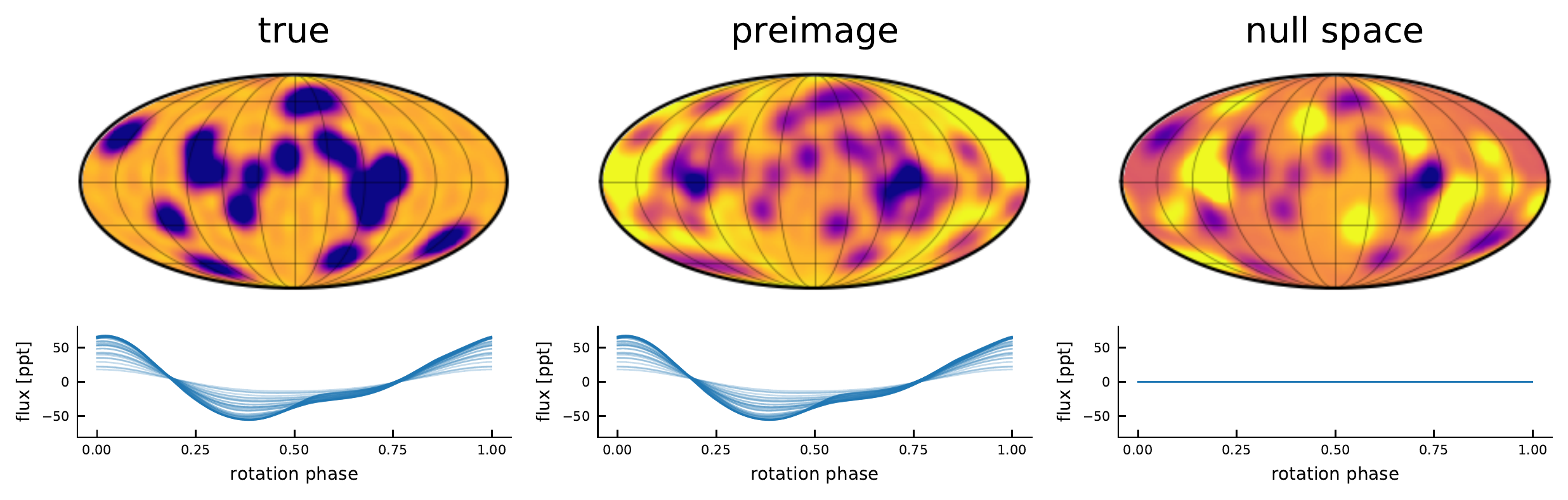}
        \\[1em]
        \includegraphics[width=\linewidth]{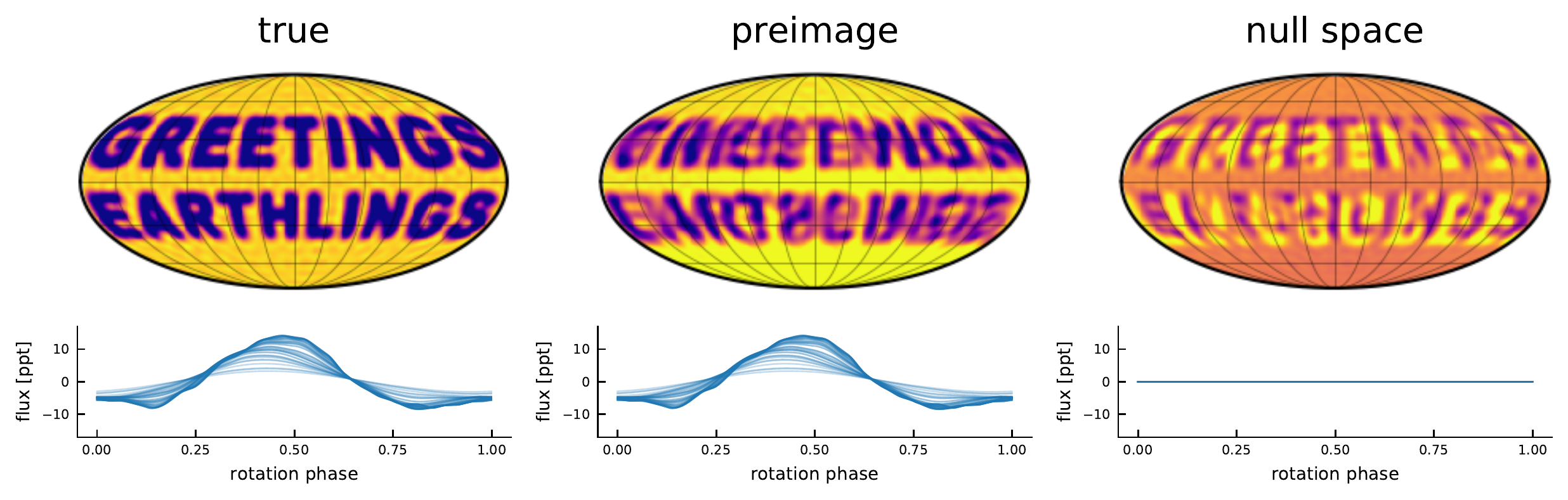}
        \oscaption{nullspace_preimage_ensemble}{%
            Same as Figures~\ref{fig:nullspace_preimage} and \ref{fig:nullspace_preimage_85}, \
            but assuming we can measure the light curves of these stars from many
            different inclinations. In this limit, the information content of our
            data approaches $50\%$ of the spatial information about the surface.
            \label{fig:nullspace_preimage_ensemble}
        }
    \end{centering}
\end{figure}

To understand what we can learn about the surfaces in this limit, let us
return to the stellar surfaces we considered in Figures~\ref{fig:nullspace_preimage}
and \ref{fig:nullspace_preimage_85}.
Figure~\ref{fig:nullspace_preimage_ensemble} shows the same decomposition of
these surfaces into preimage and null space, but this time assuming we measure
the light curves of these stars from many different random inclinations.
As expected, the partition of information between the preimage and the nullspace
is about $50{-}50$.
Because spots are compact features, they
are necessarily made up of a continuum of spherical harmonic modes spanning
many different values of $l$: they can therefore be seen in both the even
modes (the preimage) and the odd modes (the null space). The absence of
information about the odd modes therefore does not affect our ability to
infer the shape and location of the features on the surface.%
\footnote{
    There is even some hope of deciphering complex alien messages
    (last row in the figure) in this limit!
}
Interestingly, however,
the symmetries at play require spots to be paired with antipodal dark mirror
images in the preimage, and with \emph{bright} ones in the null space
(which sum to perfectly cancel out in the true map). Thus, there are still
degeneracies in this very idealized ensemble problem, but they are much easier to break
with a suitable choice of prior. For instance, in the single spot case
(top row), the ``ghost'' image in the southern hemisphere is surrounded by
a bright ring (whose effect is to cancel out its contribution to the flux);
either a compactness prior or a prior that enforces uniformity in the
background could easily penalize that feature in the fit. This may be
much harder to do for surfaces like that shown in the second row, but
we can still (in principle) learn about the size, shape, and latitude
(if not the number) of starspots from the light curves.

There are several practical reasons why this kind of ensemble analysis
may be very difficult (but not impossible!) in practice. Before
we address those, let us consider an important effect that we have not
addressed thus far in this paper: limb darkening.

\subsection{The effect of limb darkening}
\label{sec:limbdark}

\begin{figure}[t!]
    \begin{centering}
        \includegraphics[width=\linewidth]{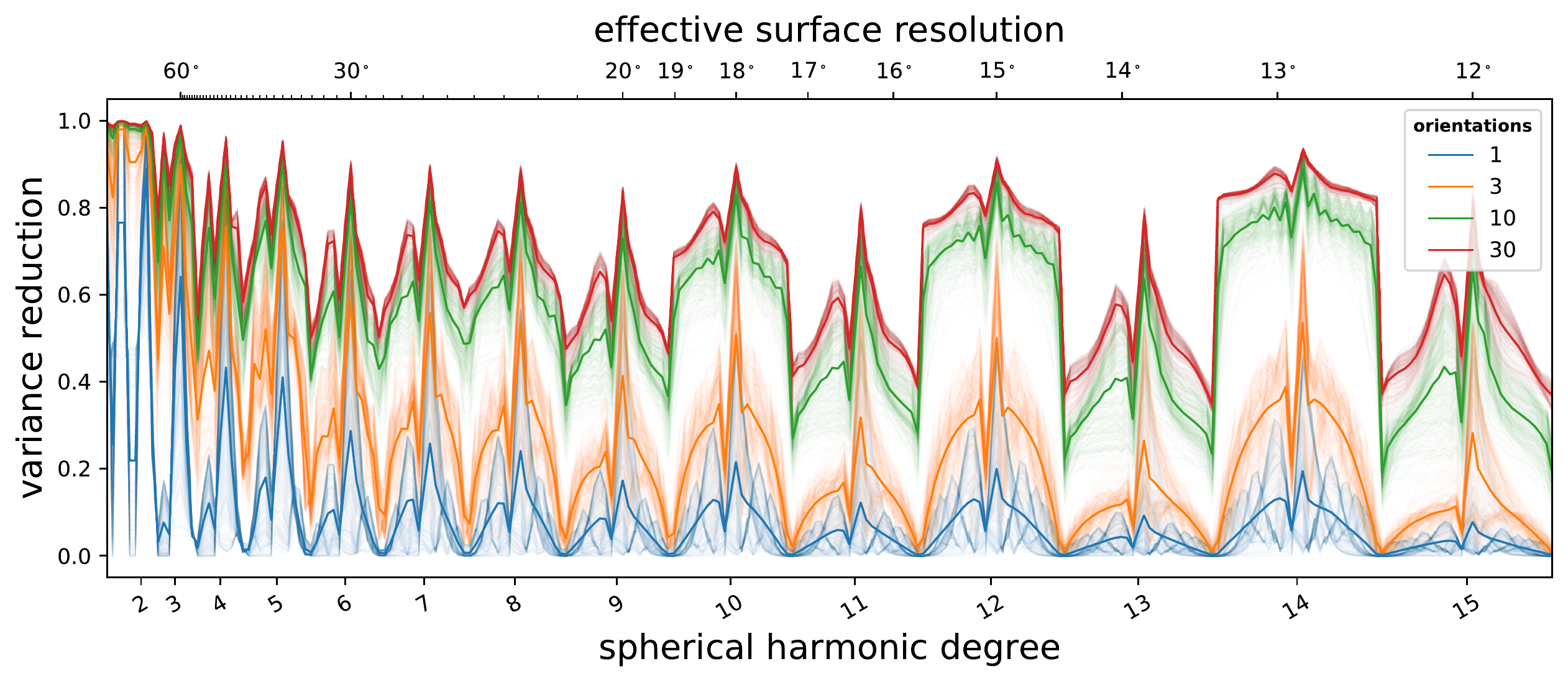}
        \oscaption{nullspace_ensemble_ld}{%
            Same as Figure~\ref{fig:nullspace_ensemble},
            but for limb-darkened
            stars with quadratic coefficients $u_1 = 0.5$ and $u_2 = 0.25$. Odd
            modes can now be probed, at the expense of the even modes.
            \label{fig:nullspace_ensemble_ld}
        }
    \end{centering}
\end{figure}

Typically, the shallower the angle between the line of sight
and the stellar surface normal, the higher up in the stellar
atmosphere optical depth unity is reached. At optical wavelengths,
lines of sight directed toward the limb of the star therefore
probe cooler temperatures, resulting in the well-known effect of
\emph{limb darkening}. Features close to the limb of the star therefore
contribute less to the total outgoing flux, and this must be
accounted for when computing the effect of a rotating starspot
on the light curve. Limb darkening is often parametrized
as a low-order polynomial in the cosine of the line of sight angle
\citep{Kopal1950}.

To understand how limb darkening affects our ability to infer
surface properties from stellar light curves,
let us repeat the experiment from the previous section,
this time with moderate quadratic limb darkening (with coefficients
$u_1 = 0.5$ and $u_2 = 0.25$, although our conclusions do not
qualitatively change for different values).
The top panel of Figure~\ref{fig:nullspace_ensemble_ld} shows the
\shrinkage plot (same as Figure~\ref{fig:nullspace_ensemble}, but
this time accounting for limb darkening). Interestingly, there is no longer a clean division of
the null space between even and odd modes in the limit of a large number
of light curves.
This is because limb darkening effectively lifts odd modes out of the null space, \emph{at the
    expense of the even modes}. While no coefficient lies entirely in the
null space ($S = 0$) when limb darkening is present, no coefficient
can be uniquely inferred ($S = 1$), either. This can be understood by noting that
a polynomial limb darkening law can be written
exactly as a linear combination of the $m=0$ spherical harmonics
up to a degree equal to the order of the limb darkening
\citep[in this case, $l = 2$;]{Luger2019,Agol2020}.
Since the limb darkening operation is a (multiplicative) downweighting of
the surface intensity, the map seen by the observer is just the product
of the spherical harmonic representation of the surface ($\mathbf{y}$)
and the spherical harmonic representation of the limb darkening profile.
And since spherical harmonics are just polynomials on the surface of the sphere,
the product of spherical harmonics of degree $l_1$ and $l_2$ is a spherical
harmonic of degree $l_1 + l_2$. This means that the linear limb darkening
component ($l = 1$) effectively raises the degree of all spherical harmonic
coefficients of the surface map by one. This has the effect of reversing
the null space: under \emph{only} linear limb darkening, it is the \emph{even}
modes that would be in the null space. However, the quadratic
limb darkening term ($l = 2$) raises the degree of all spherical harmonics by two,
so its presence ensures that the even modes can still be probed to some extent.
In reality, the true limb darkening profile of a stellar surface is more
complicated than a two-parameter quadratic model can capture; but one may still
expand it as an arbitrary order polynomial, in which case the argument still
applies---limb darkening mixes the null space and the preimage in a nontrivial
way.
The fact that no coefficient can be determined uniquely---i.e., there are
perfect degeneracies involving \emph{all} modes on the surface---could make it more
difficult in practice to perform ensemble analyses on limb-darkened stars.

\begin{figure}[t!]
    \begin{centering}
        \includegraphics[width=\linewidth]{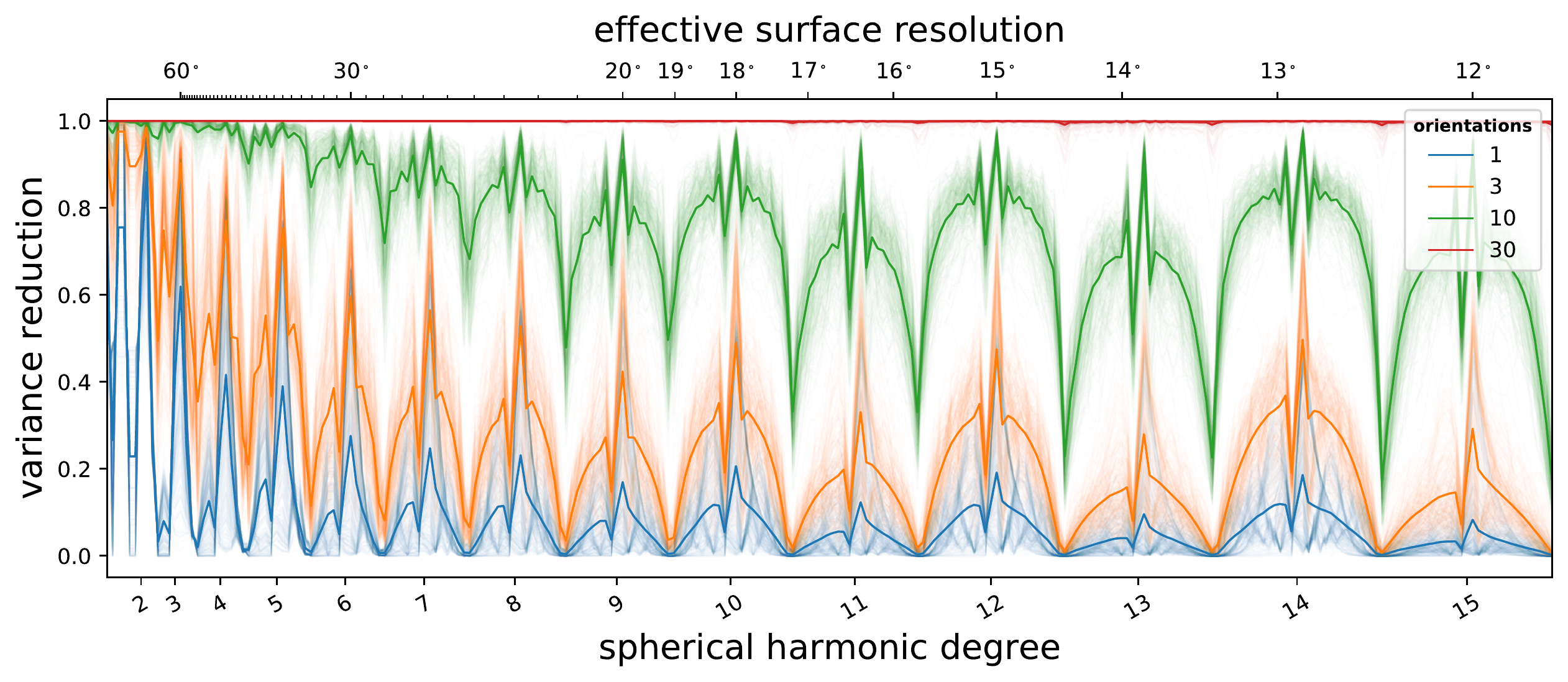}
        \oscaption{nullspace_ensemble_ld_var}{%
            Same as Figure~\ref{fig:nullspace_ensemble_ld}, but allowing for
            a 10\% variation in the limb darkening coefficients $u_1$ and $u_2$
            across different observations. For $\gtrsim 30$ light curves,
            there is virtually no null space up to at least $l_\mathrm{max} = 15$.
            \label{fig:nullspace_ensemble_ld_var}
        }
    \end{centering}
\end{figure}

In reality, it is unlikely that all stars in a given ensemble will have exactly
the same limb darkening coefficients, however ``similar'' the stars may be.
Figure~\ref{fig:nullspace_ensemble_ld_var} shows the same
\shrinkage plot as Figure~\ref{fig:nullspace_ensemble_ld},
but for the case where each star has coefficients
$u_1 = 0.5 \pm 0.05$ and $u_2 = 0.25 \pm 0.025$; i.e., we add a scatter of
10\% in the value of these coefficients. The plot shows the \shrinkage
in the hypothetical case where we know the coefficients for
each star exactly. Now, as the size of the ensemble increases, the \shrinkage
approaches unity \emph{for all spherical harmonic modes}.
In the same way that the null space is a strong function of the inclination,
allowing us to chip away at it with observations at different inclinations,
the null space is also a strong function of the limb darkening law. Even a
small amount of variance in the coefficients is sufficient to constrain all surface
modes exactly (in the limit of infinite SNR and a large number of light curves).

In practice, of course, we will never know the limb darkening coefficients
exactly. In the next section, we will revisit this and other assumptions we
made above in a more sober light.

\subsection{A reality check}
\label{sec:reality-check}

There are three major points that make the kind of ensemble
analyses discussed above difficult in practice. First, and perhaps
most obviously, stars are
not identical, no matter how ``similar'' we think they may be.
Even stars of the same spectral type, age, and metallicity will
in general have different configurations of spots on their surfaces.
When we perform an ensemble analysis on the light curves of a
heterogeneous group of stars, we are learning something about the
\emph{distribution} of surface properties across all the stars in
the sample---not the surface properties of any individual star.
What exactly we can learn in this case is not immediately obvious,
and requires a detailed investigation. This is the subject of
the next paper in this series \citepalias{PaperII}, where we show that,
armed with a good model,
we can learn \emph{a lot} about the distribution of starspot
properties of a heterogenous group of stellar light curves.

The second point is that while observing many similar stars at different
inclinations can greatly help us learn about their surfaces from a
statistical standpoint, our analysis
above assumed we knew what the values of the individual inclinations were.
In practice, this will usually not be the case. While we may have
good priors for some stars (from spectroscopic $v\sin I$ measurements,
or from the assumption that transiting exoplanet hosts are likely
to have inclinations close to $I=90^\circ$), for the vast majority
of field stars we won't know much a priori. Since the inclination
is typically degenerate with the spot latitude \citep[e.g.,][]{Walkowicz2013},
this decreases the constraining power of ensemble analyses.
However, as we show in \citetalias{PaperII}, there is still enough
information in ensembles of $\gtrsim 50$ light curves
to independently constrain the spot latitudes
and the \emph{individual} stellar inclinations.

The final point concerns limb darkening, which also has a strong
effect on the structure of the null space. While limb darkening can
help us in the same way as the inclination, in practice it is likely to be more
of a problem, since the use of incorrect limb darkening coefficients can lead
to bias in the spot properties when doing inference.
It is therefore extremely important to use reliable limb darkening models
when doing ensemble analyses; we also explore this in \citetalias{PaperII}.

\section{The normalization degeneracy}
\label{sec:normalization}

Thus far we have focused our discussion on theoretical aspects concerning
what can and cannot be learned from disk-integrated photometric measurements
of stellar surfaces. In this section, we discuss an important degeneracy
introduced by how we actually measure stellar light curves, which we
will refer to as the \emph{normalization degeneracy}. This degeneracy has been
pointed out before \citep[e.g.,][]{Basri2018}, but it is useful to revisit and
build on it here.

\subsection{A fundamental issue of units}
To understand the normalization degeneracy, consider how we might go about simulating
a stellar surface.
We might add a dark spot somewhere on the surface, either by expanding
it in spherical harmonics or by gridding up the stellar surface and
setting the intensity of pixels within the spot to a low value. To compute
the light curve, we integrate over the
projected disk at each point in time.
The resulting
light curve will have strange units, so we might then divide by the
integral of the background intensity over the unit disk, so that we are
now in what we will call \emph{fractional units}: the flux as a
fraction of the flux we would measure if the star had no spots on it.

\begin{figure}[t!]
    \begin{centering}
        \includegraphics[width=\linewidth]{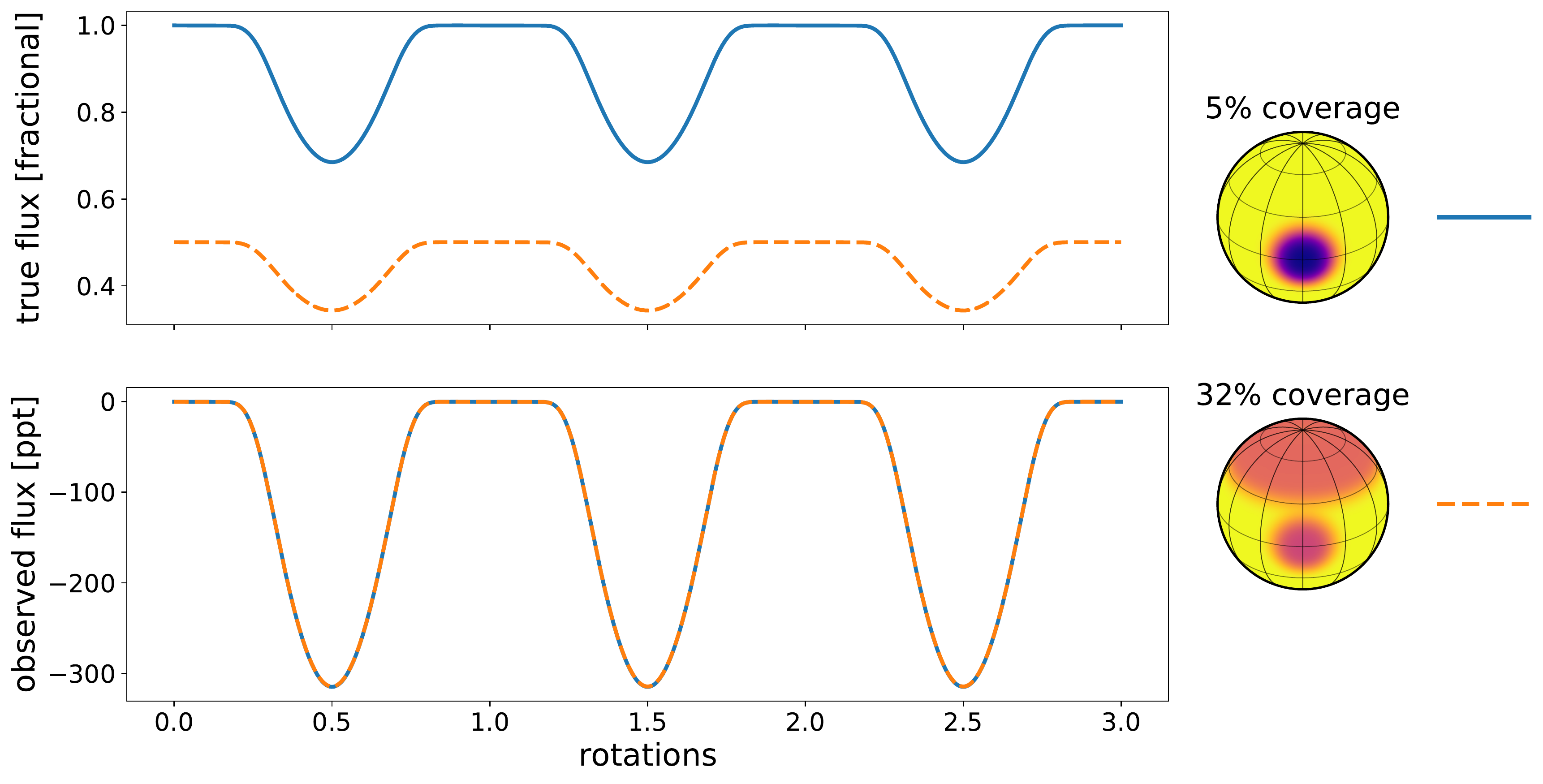}
        \oscaption{mean_normalization}{%
            An example of the normalization problem.
            \emph{Top:} Consider a star with a single
            equatorial spot of contrast $c$ viewed at a certain inclination.
            The total flux (in some units) as a function of time is shown as
            the blue curve. Now, consider a second star,
            identical in all respects to the first, except that (1) the equatorial spot
            has half the contrast (i.e., $\nicefrac{c}{2}$); and (2) there is
            a second, large spot centered on the pole. The corresponding light
            curve is shown as the dashed orange curve.
            The orange light curve is different from the blue one in two ways:
            (1) since the equatorial spot has half the contrast, the amplitude of the associated
            dips in the light curve is half that of the first star; and
            (2) since the polar spot is azimuthally symmetric, its only
            contribution is a net darkening at all phases.
            \emph{Bottom:} The true baseline level of a stellar
            light curve, which corresponds to the flux one would measure in the
            absence of any spots, is almost always unknown. Photometric measurements
            are therefore meaningful only in a relative sense, i.e., as deviations from
            the mean, median, or maximum level of the light curve.
            The bottom panel shows the same two light curves, this time plotted as
            deviations in parts per thousand (ppt) from their respective maxima.
            To the observer, the two light curves are \emph{indistinguishable}.
            In the absence of baseline information, there
            exists a perfect degeneracy between the total spot coverage
            and the contrast of any individual feature on the surface.
            As a consequence, the total spot coverage of a star cannot be
            uniquely inferred from single-band photometry.
            \label{fig:mean_normalization}
        }
    \end{centering}
\end{figure}

The top panel of Figure~\ref{fig:mean_normalization} shows two mock
light curves we might compute following the procedure above. The solid blue
curve corresponds to the light curve of a star with a single large
equatorial spot of contrast $c$ viewed at an inclination $I = 60^\circ$.
The dashed orange curve corresponds to a star with a spot at the same
location but half the contrast, plus a large polar spot of comparable
contrast. Because the equatorial spot on this star has half the contrast of that on
the first star, the peak-to-trough amplitude of the orange light curve is half that of
the blue light curve. Moreover, since the polar spot is always in view on
this star, the peak flux is itself only about half that of the first star.
If we were given these two light curves \emph{in these fractional units},
we might be able to infer these basic differences between the two stars
(setting aside for the moment all the issues with the null space discussed
in the previous section).

However, \textbf{we do not observe stellar light curves in fractional units.}
Instead, we typically observe in units of counts on
the detector, which depend (among other things) on various properties of the telescope.
But even if we knew all these things, as well as the true luminosity
of the star and the precise distance to it, and we could truly
perform \emph{absolute photometry}, that \emph{still would not be enough}
to correctly calibrate the light curve into the units we need.
\textbf{To convert to fractional units, we would need to know the brightness we would
    measure if the star had no spots on it.} This depends on the brightness
of the unspotted photosphere, which cannot be measured directly unless
we actually resolve the star (i.e., interferometrically)!

Usually, this isn't much of an issue. Astronomers typically circumvent
this by self-normalizing the data: i.e., dividing the flux by the mean,
median, maximum, or some similar statistic of the light curve.
This operation folds the unknowability of
the true units under the rug and transforms the light curve into a \emph{relative}
measurement of the star's temporal variability. While relative measurements
are typically what we are interested in anyways, this normalization procedure
can sometimes fool us into thinking we have access to information that
is simply not observable in single-band photometry. To understand why,
consider the lower panel of Figure~\ref{fig:mean_normalization}, which shows the
same two light curves in what we will call \emph{relative units}. To
get the light curves in these units, we followed the common procedure of
dividing each by the observed ``continuum'' level (the maximum flux in the
light curve), subtracting unity, and multiplying by $1{,}000$, yielding
relative fluxes in units of parts per thousand (ppt).

The two light curves, which were distinct in the fractional units we used
to generate them, are \textbf{indistinguishable in the relative units
    in which we observe them}. There is absolutely no information in the
relative light curves that can differentiate between the two stellar
surfaces shown in the figure. The depth of each of the dips cannot tell
us about either the contrast of individual spots or the total number of spots;
in fact, in single-band photometry, these two quantities are perfectly
degenerate with each other.

\subsection{Ensemble analyses don't necessarily help}
\label{sec:basic-gp}
To explore this point in a bit more detail, let us consider the normalization
degeneracy in the context of ensemble analyses. Even though we can't
uniquely infer contrasts or numbers of spots from individual light curves,
perhaps we could harness the power of the ensemble.
Let us therefore go
back to our thought experiment in which we added spots to a stellar
surface. Assuming for simplicity that all spots have the same contrast $c$,
every time we add a spot the flux (in fractional units)
decreases by an amount proportional to $c$, so to first order we can approximate an
individual light curve as
\begin{align}
    \label{eq:fapprox}
    \mathbf{f}(c, n) = \mathbf{1} - c \sum_{i=0}^{n-1} \mathbf{g}(\pmb{\theta}_i)
\end{align}
where $\mathbf{g}(\pmb{\theta}_i)$ is some (complicated) function of the properties
of the $i^\mathrm{th}$ spot
as well as the stellar inclination, rotation period, etc., which we
denote by $\pmb{\theta}_i$.
Now, consider many stellar light curves drawn from some distribution
controlling the stellar and starspot properties with probability density
$p(\pmb{\theta})$.
The mean of the distribution of
light curves (still in fractional units) is then given by
\begin{align}
    \mu(c, n) & = \mathrm{E} \Big[ \mathbf{f} (c, n) \Big] \nonumber                                      \\
              & = 1 - c \, \mathrm{E} \bigg[ \sum_{i=0}^{n-1} \mathbf{g}(\pmb{\theta}_i) \bigg] \nonumber \\
              & = 1 - c \, n \, \alpha
    \quad,
\end{align}
where $\mathrm{E}\big[\cdots\big]$ denotes the expected value and
\begin{align}
    \alpha \equiv \int \mathbf{g} (\pmb{\theta}) p(\pmb{\theta}) \mathrm{d} \pmb{\theta}
\end{align}
is the expected value of $\mathbf{g}$.
Similarly, the variance of the distribution may be computed as
\begin{align}
    \sigma^2(c, n) & = \mathrm{Var} \Big[ \mathbf{f}(c, n) \Big] \nonumber                         \nonumber \\
                   & = \mathrm{Var} \Big[ c \, \sum_{i=0}^{n-1} \mathbf{g}(\pmb{\theta}_i) \Big]   \nonumber \\
                   & = c^2 \, \mathrm{Var} \Big[ \sum_{i=0}^{n-1} \mathbf{g}(\pmb{\theta}_i) \Big] \nonumber \\
                   & = c^2 \sum_{i=0}^{n-1} \mathrm{Var} \Big[  \mathbf{g}(\pmb{\theta}_i) \Big]   \nonumber \\
                   & = c^2 \, n \, (\beta^2 - \alpha^2)
    \quad,
\end{align}
where $\mathrm{Var}\big[\cdots\big]$ denotes the variance,
\begin{align}
    \beta^2 \equiv \int \mathbf{g}^2 (\pmb{\theta}) p(\pmb{\theta}) \mathrm{d} \pmb{\theta}
\end{align}
is the expected value of $\mathbf{g}^2$, and we used the fact that the variance
of the sum of independent random variables is equal to the sum of their variances.

To summarize, the mean and variance of the ensemble of stellar light curves
in fractional units is
\begin{align}
    \mu      & = 1 - c \, n \, \alpha
    \nonumber                                     \\
    \sigma^2 & = c^2 \, n \, (\beta^2 - \alpha^2)
    \quad,
\end{align}
for some complicated functions $\alpha$ and $\beta$ of the distribution of
stellar inclinations,
rotation periods, and starspot properties.
If our observations were collected in these fractional units,
we could uniquely infer the spot contrast $c$ and the number of spots $n$,
since these enter as $c \, n$ and $c^2 n$ in the expressions for the mean
and variance, respectively, and these are straightforward statistics to
compute from the ensemble.
However, because of observations are made in \emph{relative} units, in which
we typically normalize to the mean, the amplitudes of features in light curves
can only tell us about the \emph{ratio}
\begin{align}
    \label{eq:ratio}
    \frac{\sigma}{\mu}
     & \propto \frac{c \sqrt{n}}{1 - c n \alpha}
\end{align}
for some value of $\alpha$.
In other words, even photometric ensemble analyses may not be able to tell
us about the values of the contrast and the number of spots independently.

A direct consequence of this normalization degeneracy is that it may not be
possible to uniquely constrain the total spot coverage of a star from
single-band photometry without strong prior assumptions.
The total spot coverage $f_S$
is simply the (average) area of a spot
times the total number of spots divided by the total area of the sphere, which
may be expressed as
\begin{proof}{test_fS}
    \label{eq:fS}
    f_S = \frac{1}{2}\left(1 - \left<\cos r\right>\right)n
    \quad,
\end{proof}
given an angular spot radius $r$.
While $r$ may be uniquely constrained from the covariance structure of the data
\citepalias{PaperII}, $n$ cannot.

The arguments above are heuristic and based on only the first two moments
of the distribution of light curves in an ensemble.
It is possible, at least in principle, that higher order moments of the data
could encode information to
break the $c-n$ degeneracy, but these are in general more difficult
to constrain from the data. It is also possible that we could place a
\emph{lower limit} on the spot contrast based on the normalized light
curve. In Equation~(\ref{eq:fapprox}) we implicitly assumed that
spots are allowed to overlap; under this assumption, it is possible to
double the contrast of a spot by simply adding another spot on top
of it. We could therefore generate light curves with arbitrarily
large dips by choosing a small value for $c$ and a very large value for $n$.
In reality, spots do not behave in this way: many overlapping spots
likely still have the same effective temperature! Therefore,
assuming there are no bright spots,
the maximum depth of a feature in the light curve (even in relative units)
could be used to place a lower limit on the spot contrast.
Moreover, in \citetalias{PaperII} we empirically show that, depending on the choice
of prior, ensemble analyses may be able to place stronger constraints on
the contrast $c$ than on the number of spots $n$.

Nevertheless, even with these caveats in mind, the fact remains
that degeneracies like the polar spot effect are fundamental: recall
Figure~\ref{fig:mean_normalization}, in which the two stars
have identical relative light curves, but very different spot coverage
fractions. In general, any azimuthally-symmetric mode on the surface
is in the null space of the \emph{normalized} light curve problem. These
components contribute a constant value to the flux at all phases, which
effectively gets normalized away when we measure the light curve. A
polar spot is just one manifestation of this degeneracy. Bands, or
band-like arrangements of spots, will be at least partly in the null
space of normalized light curves. These features change the total ``spot coverage''
of the surface but do not affect the light curve in a uniquely measurable
way.

\subsection{Effect on the covariance structure}
\label{sec:covariance}

There is one final subtle point concerning the normalization degeneracy that merits discussion.
The common procedure of normalizing light curves to their mean, median, or
maximum level does not only change the \emph{units} of the data: it changes
the very covariance structure of the light curves.

To understand why, let us consider the procedure of normalizing a light curve
to its mean value. Whenever we scale our data, we must always be sure to
scale the errorbars accordingly. Since in this case we are dividing the
flux by the mean, one might imagine that we could simply divide each of
the measurement uncertainties by the same amount. However,
this is technically incorrect!

\begin{figure}[t!]
    \begin{centering}
        \includegraphics[width=\linewidth]{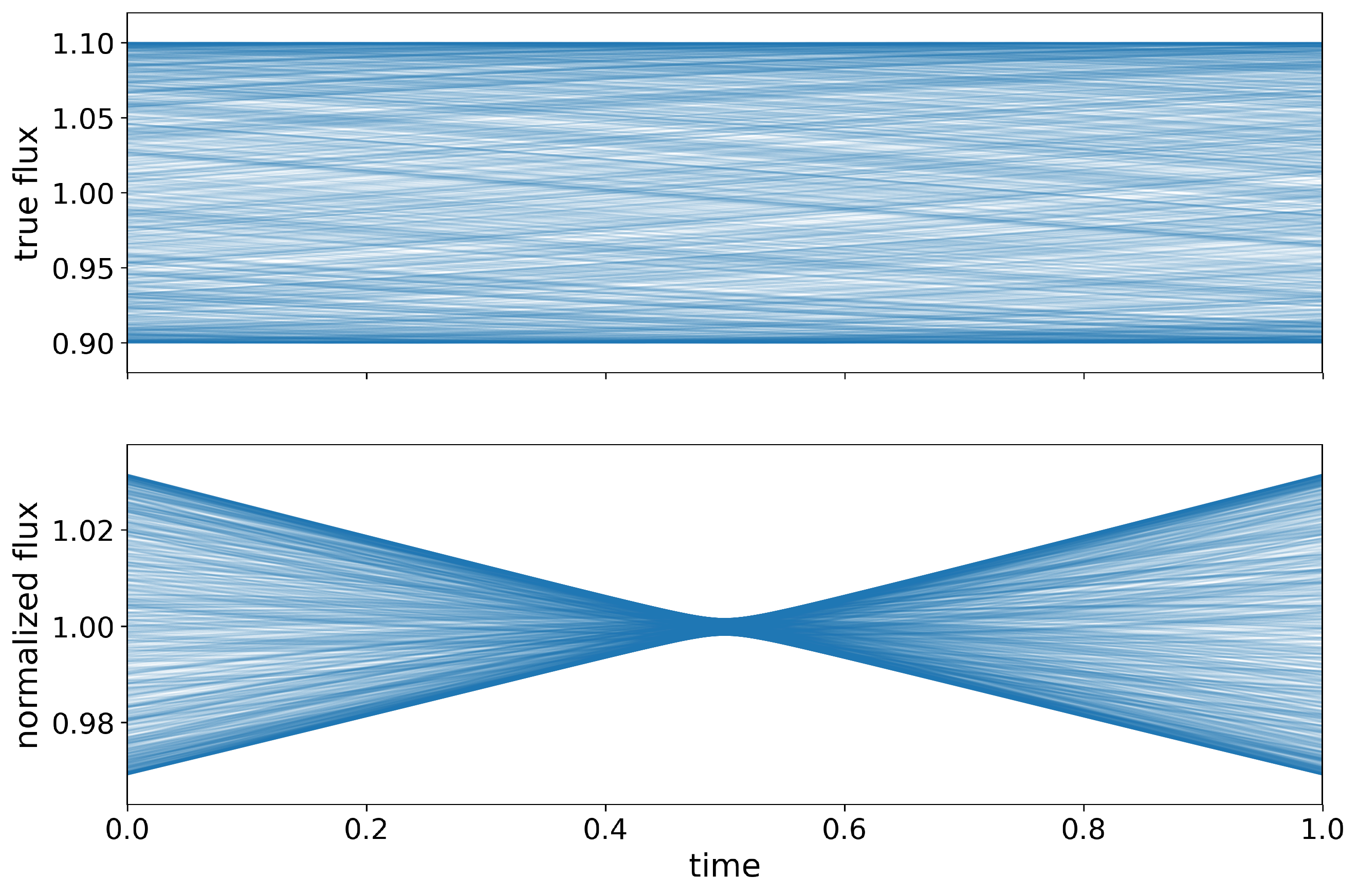}
        \oscaption{nonstationarity}{%
            An example of why normalized light curves are non-stationary.
            The top panel shows $1{,}000$ samples from a unit-mean sinusoid with
            an amplitude of 10\% and a period of 10 days, much longer than the
            1 day observation baseline. The bottom panel shows the same light curves,
            each normalized to its own mean. Because the mean tends to be near
            the center of the observation window, points near $t=0.5$ are driven
            to values very close to unity, while points near the edges have much
            larger scatter.
            \label{fig:nonstationarity}
        }
    \end{centering}
\end{figure}

Consider the example in Figure~\ref{fig:nonstationarity}. The top panel
shows $1{,}000$ samples from a sinusoid with random phases and a period
equal to ten times the observational window.
In this limit, each light curve is approximately
linear, which causes its mean value to roughly coincide with the midpoint of the
observation window.
Division by the mean value (lower panel) results in points near the
midpoint being driven to unity and points near the edges (whose values differ
the most from the mean) to be driven to both large and small values.
If our error bars in the original data were uniform (homoscedastic),
the error bars in the normalized light curves are
not: the standard deviation (or variance) of the data is now
distinctly dependent on the phase.

While the example shown in the figure is fairly extreme, the idea here
is quite general: the normalization procedure changes the covariance
structure of the data. In most cases, the non-stationarity (i.e., the
phase dependence) of the variance will be quite small. The effect is primarily
important for light curves with periods much longer than
the observation window. In these cases, not accounting for this effect
could introduce bias in light curve analyses. A detailed investigation of
this effect is beyond the scope of this paper, but we do present a
method to correct the covariance matrix of normalized light curves for
this issue in \citetalias{PaperII}.

\section{Discussion and conclusions}

\subsection{Degeneracies stemming from the null space}

In this paper we explored various degeneracies in the problem of
inferring a stellar surface map from its rotational light curve.
We discussed the idea behind the null space, the set of surface modes that
have exactly zero effect on the observed light curve (\S\ref{sec:nullspace}).
For rotational light curves, we showed that the vast majority of the information
about the surface intensity is in the null space and therefore cannot
be inferred from unresolved photometric measurements. We showed, in particular,
that the size of the null space grows \emph{quadratically} with increasing spatial
resolution, while the number of independent degrees of freedom in the light curve
only grow \emph{linearly}. Consequently, the information content of light
curves is small for large scale surface features and \emph{vanishingly small}
for small scale features.

A direct consequence of this fact is that \textbf{results based on analyses
    of individual light curves are often extremely sensitive to
    the particular assumptions of one's model.} We therefore urge lots of care
in these kinds of analyses, where assumptions of uniform-contrast, discrete, and
circular spots, or alternatively of maximum entropy or ``simplicity'' may
result in significant bias. In this respect, we agree with the recent
paper by \citet{Basri2020}, which cautions against the association of
invidual dips in light curves with individual, discrete spots.

One of the main results in this paper concerning the null space is its
dependence on stellar inclination (\S\ref{sec:inclination}).
Because the modes that lie in the null
space change depending on the viewing angle, observations of a star at many
inclinations would break many of the degeneracies in the mapping problem. While
this is obviously infeasible in practice, the dependence of the null
space on inclination motivates ensemble analyses of many similar stars
as a way of circumventing the mapping degeneracies by
providing a strong data-driven prior. As we show in
\citetalias{PaperII} in this series, the joint analysis of tens to hundreds
of light curves can uniquely constrain the distribution of starspot
sizes and latitudes among the stars in the sample. In that paper,
we show that ensemble analyses can
even break the latitude-inclination degeneracy \citep[e.g.,][]{Walkowicz2013},
allowing one to infer individual stellar inclinations, typically to within
about $10^\circ$.

We also showed how the null space is a strong function of the stellar limb
darkening. While this can again be used to our advantage---by harnessing the
variance in the limb darkening parameters across an ensemble of stellar
light curves---in practice it is likely to complicate inference, since
any bias in the assumed limb darkening coefficients will likely result in
bias in the inferred surface properties. We revisit this point in
\citetalias{PaperII}.

One final point that we did not address thus far concerns occultations.
Our results regarding the null space apply strictly to rotational light curves,
in which all points on the projected disk contribute to the measured flux.
During an occultation by a transiting planet or a binary companion, the
weighting of surface modes giving rise to the light curve changes
substantially. In fact, the presence of an occultor breaks many of the perfect
symmetries that give rise to a null space in the first place
\citep[e.g.,][]{Luger2019}. This fact can be used to infer the properties
of spots in the path of the occultor, as was done (for example)
by \citet[e.g.,][]{Morris2017}. A detailed investigation of the null space
for the occultation problem is beyond the scope of this work.

\subsection{Degeneracies due to the unknown normalization}

The second major source of degeneracies is the fundamental
unknowability of the true normalization in single-band
photometry (\S\ref{sec:normalization}), which is summarized in
Figure~\ref{fig:mean_normalization}. In a nutshell, the relationship
between the amplitude of a feature in the light curve and the contrast
of the feature that gave rise to it depends on quantities like the
unspotted photospheric brightness, which is not an observable in
single-band photometry. This leads to the possibility of distinctly
different stellar surfaces having identical \emph{relative}
light curves, as demonstrated in the figure.

In practice, this degeneracy usually manifests as a nonlinear correlation
between the contrast $c$ of a spot and the total number $n$
of spots on the surface of the star. Even in ensemble analyses,
one cannot in general learn these two quantities independently from
the data alone,
only a (complicated) function of the two (Equation~\ref{eq:ratio}).
In \citetalias{PaperII} we show empirically that
careful analysis of the covariance structure of ensembles of light curves
may shed some light on $c$ but cannot uniquely constrain $n$.
A direct consequence of this degeneracy is that \textbf{constraints on
    the total number of spots or on the spot coverage of a star from single-band light curves
    depend strongly on the model assumptions.}

Recently, \citet{Morris2020} performed an ensemble analysis of
\emph{Kepler}, \emph{K2}, and \emph{TESS}
light curves to derive a relationship between the fractional spot
coverage $f_S$ and stellar age.
%
That work found that the spot coverage $f_S$ as a
function of age is well modeled by a simple power law, decreasing from
${\sim}10\%$ for the youngest (${\sim}10$ Myr) stars
to less than $1\%$ for the oldest (${\sim}5$ Gyr) stars.
While this broadly agrees with the expectation that stellar
magnetic activity decreases over time, our work strongly suggests
that these results depend heavily on the prior.
This is because the expression for $f_S$ (Equation~\ref{eq:fS}) depends on
two quantities: the average spot radius $r$, which can be
constrained \citep[see][]{PaperII}, and the total number of spots $n$, which
we showed \emph{cannot} be uniquely constrained from single-band light curves.
In fact, \citet{Morris2020} assumed $n=3$ for simplicity when doing
posterior inference. We therefore urge care in interpreting those results,
as it is possible this assumption does not hold across the large
range of spectral types and stellar ages considered in that study.
Nevertheless, the other central result of that paper, the relationship
between the ``smoothed amplitude'' of a light curve and the stellar age,
\emph{is} valid.
Moreover, the core idea in \citet{Morris2020}
\citep[and in related studies such as][]{Jackson2013}
is very similar
to that advocated here: the use of ensemble analyses to constrain population-level
parameters when invidual datasets are not sufficiently constraining.

Another recent paper relevant to our work is that of \citet{Basri2020}, who
investigated the information content of stellar light curves, exploring what
can and cannot be learned about star spot configurations from individual
light curves. That paper strongly urges against the common practice
of interpreting
light curves with one or two dips as originating from one or two spots,
respectively, a point we strongly agree with (see our Figure~\ref{fig:degeneracies}).
It also reinforces our point about the additional degeneracies introduced
by the unknown normalization inherent to single-band photometry.
%


As with the degeneracies due to the null space, there are potential ways
to break the normalization degeneracy.
In principle, the maximum level of a light curve could set a lower limit on
the brightness of the
unspotted photosphere, particularly for long-baseline, time-variable
surfaces \citep{Basri2018}. However, this would work only if the surface is
known to be made up \emph{exclusively} of dark spots. The presence of bright
spots (faculae), which are common on the Sun, make it difficult
for one to infer this quantity (and hence the correct normalization)
from single-band photometry in practice.

A better approach may be to collect photometric data in
multiple wavelength bands, an idea that has
been explored recently \citep[e.g.,][]{Gully2017,Guo2018}. Assuming the
locations and sizes of a star's spots are constant in wavelength, the amplitude
of the light curve in different bands (and in particular its slope as a function
of wavelength) can be used to directly constrain the
temperature, and hence the contrast, of the spots. This effectively breaks the
$c-n$ degeneracy. In practice, the effective size of spots may be different
at different wavelengths, which could complicate this picture somewhat, but
the point still stands that light curves collected in multiple bands
contain at least \emph{partial} information about the correct normalization.
A more detailed analysis of the information content of multi-band light curves,
and, by extension, spectroscopic timeseries, is deferred to a future paper
in this series.

\subsection{Caveats}
\label{sec:caveats}

There are several simplifying assumptions we made in this paper that
are worth discussing. First, in our characterization of the null space
we assumed we knew the stellar inclination, the stellar rotation period,
and the limb darkening coefficients
exactly, and we computed the information content of stellar light curves
in the limit of infinite SNR. None of these assumptions are valid in
practice. In realistic scenarios, the \shrinkage will be
necessarily \emph{lower} than what we presented here: the curves in
Figure~\ref{fig:nullspace_ensemble}, for instance, are therefore
strict upper bounds on the amount of information that can be learned about
the surface. In some cases, it may be possible to infer the inclination
from spectroscopic $v \sin I$ measurements, or the limb darkening coefficients
from the shape of transiting exoplanet light curves; but any uncertainty
in these quantities will degrade our ability to constrain
information about the surface.

Moreover, our constraints on the information content of ensembles of light curves are
also strict upper limits, since we assumed all stars in the ensemble
are identical. Variance in the surface maps of stars in
a population makes it effectively impossible to infer detailed
properties of the individual stellar surfaces in the ensemble.
We argued without proof that what we can learn instead are properties of the
\emph{distribution} of stellar surfaces among the population, such as
the distribution of spot sizes and active latitudes.
Demonstrating this point is more difficult, so we defer it to
\citetalias{PaperII}, where we construct a custom Gaussian process model
and show that it can be used to infer such population-level parameters
from ensembles of light curves.

Finally, we limited our discussion to static stellar surfaces.
Real stellar surfaces vary
with time as individual spots form, evolve, and dissipate, or as a star progresses
through its activity cycle. This, too, makes it more difficult to constrain
the stellar surface at a particular point in time, since the amount of
data corresponding to a particular surface configuration is more limited.
On the other hand, a time-variable surface could be used to our advantage
from an ensemble standpoint. Although we are unable to uniquely infer
invididual spot properties, we could treat each point in the light curve
as a realization of a random process (i.e., a draw from some distribution)
describing the stellar surface at a high level. In this sense, we could
harness the fact that we can measure the light curves of many ``similar''
surfaces to learn something about their statistical properties.
Demonstrating this is beyond the scope of the present work, but we
return to this point in \citetalias{PaperII}.

\subsection{Future work}

This paper is the first in a series dedicated to developing
methodology to infer stellar surface properties from
unresolved measurements. It sets the stage for \citet{PaperII},
in which we develop an interpretable Gaussian process (GP) model for
starspot-induced light curve variability. This GP is aimed specifically
at the difficult problem of jointly modeling many light curves
in an ensemble analysis. Although we hinted at this here, we will show
in that paper that ensemble analyses can uniquely constrain several
statistical properties of starspots, including their distribution
of radii and latitudes across stars in the ensemble. In that work
we also consider temporally evolving surfaces, which we neglected
in our discussion thus far.

Finally, while this paper explicitly dealt with the problem of
photometric rotational light curves, the degeneracies outlined
here and the methodology developed in this series of papers
apply more broadly in other contexts.
These including applications where the stellar
surface is a nuisance (exoplanet detection and characterization using
transit light curves or radial velocities) and spectral time series
datasets (such as transmission spectroscopy and Doppler imaging).

\vspace{2em}

In keeping with other papers in the \starry series, all figures in this
paper are generated automatically from open-source scripts linked to in
each of the captions \codeicon, and the principal equations link to associated
unit tests that ensure the accuracy and reproducibility of the algorithm
presented here \testpassicon/\testfailicon.

\vspace{2em}

We would like to thank Michael Gully-Santiago, Fran Bartoli\'c, and the
Astronomical Data Group at the Center for Computational Astrophysics for
many thought-provoking discussions that helped shape this paper.

\bibliography{bib}

\appendix

\section{Decomposition of the flux operator}
\label{sec:app-svd}
In this section we show how to use singular value decomposition (SVD)
to decompose a surface map
into its preimage and its null space (see \S\ref{sec:svd}). By performing SVD, we
may express the flux design matrix as
\begin{align}
    \label{eq:svd}
    \pmb{\mathcal{A}} = \mathbf{U} \, \mathbf{S} \, \mathbf{V}^\top
    \quad,
\end{align}
where $\mathbf{U}$ is a $(K \times K)$ orthogonal matrix,
$\mathbf{V}$ is a $(N \times N)$ orthogonal matrix,
and $\mathbf{S}$ is a $(K \times N)$ diagonal matrix.
The columns of $\mathbf{U}$ and the rows of $\mathbf{V}^\top$ are the left and right
\emph{singular vectors} of $\pmb{\mathcal{A}}$, and the entries along the
diagonal of $\mathbf{S}$ are the corresponding \emph{singular values}, arranged
in descending order. If $\pmb{\mathcal{A}}$
has rank $R$, the first $R$ singular values will be nonzero, while the
remaining $N - R$ will be identically zero.
If we assume for definitess that $K > N$ (i.e., we have more flux observations
than surface map coefficients we're trying to constrain), we can express the
matrices in Equation~(\ref{eq:svd}) as
\begin{align}
    \mathbf{U}
     & =
    \left(
    \begin{array}{ccc|ccc}
            \mathbf{U}_{0,0}          & \cdots & \mathbf{U}_{0,R\text{-}1}          & \mathbf{U}_{0,R}          & \cdots & \mathbf{U}_{0,K\text{-}1}          \\
            \vdots                    & \cdots & \vdots                             & \vdots                    & \cdots & \vdots                             \\
            \mathbf{U}_{K\text{-}1,0} & \cdots & \mathbf{U}_{K\text{-}1,R\text{-}1} & \mathbf{U}_{K\text{-}1,R} & \cdots & \mathbf{U}_{K\text{-}1,K\text{-}1}
        \end{array}
    \right)
    \equiv
    \left(
    \begin{array}{c|c}
            \mathbf{U}_\bullet & \mathbf{U}_\circ
        \end{array}
    \right)
    \\[1.5em]
    \label{eq:S}
    \mathbf{S}
     & =
    \left(
    \begin{array}{ccc|ccc}
            \mathbf{S}_{0,0}\phantom{y} &                                     &                                    &                             &                                     &                                    \\
                                        & \ddots                              &                                    &                             & \mbox{\normalfont\Large\bfseries 0} &                                    \\
                                        &                                     & \mathbf{S}_{R\text{-}1,R\text{-}1} &                             &                                     &                                    \\
            \hline
                                        &                                     &                                    & \mathbf{S}_{R,R}\phantom{y} &                                     &                                    \\
                                        & \mbox{\normalfont\Large\bfseries 0} &                                    &                             & \ddots                              &                                    \\
                                        &                                     &                                    &                             &                                     & \mathbf{S}_{N\text{-}1,N\text{-}1} \\
                                        &                                     &                                    &                             &                                     &                                    \\
                                        & \mbox{\normalfont\Large\bfseries 0} &                                    &                             & \mbox{\normalfont\Large\bfseries 0} &                                    \\
                                        &                                     &                                    &                             &                                     &
        \end{array}
    \right)
    \equiv
    \left(
    \begin{array}{c|c}
            \mathbf{S}_\bullet & \mathbf{0}       \\
            \hline
            \mathbf{0}         & \mathbf{S}_\circ
        \end{array}
    \right)
    \\[1.5em]
    \mathbf{V}^\top
     & =
    \left(
    \begin{array}{cccccc}
            \mathbf{V}_{0,0}^\top          & \cdots & \mathbf{V}_{0,N\text{-}1}^\top          \\
            \vdots                         & \cdots & \vdots                                  \\
            \mathbf{V}_{R\text{-}1,0}^\top & \cdots & \mathbf{V}_{R\text{-}1,N\text{-}1}^\top \\[0.5em]
            \hline                                                                            \\[-0.85em]
            \mathbf{V}_{R,0}^\top          & \cdots & \mathbf{V}_{R,N\text{-}1}^\top          \\
            \vdots                         & \cdots & \vdots                                  \\
            \mathbf{V}_{N\text{-}1,0}^\top & \cdots & \mathbf{V}_{N\text{-}1,N\text{-}1}^\top
        \end{array}
    \right)
    \equiv
    \left(
    \begin{array}{cc}
            \mathbf{V}_\bullet^\top \\
            \hline
            \mathbf{V}_\circ^\top
        \end{array}
    \right)
\end{align}
where
$\mathbf{U}_\bullet$ is $(K \times R)$,
$\mathbf{U}_\circ$ is $(K \times K - R)$,
$\mathbf{S}_\bullet$ is $(R \times R)$,
$\mathbf{S}_\circ$ is $(K - R \times N - R)$,
$\mathbf{V}_\bullet^\top$ is $(R \times N)$,
and
$\mathbf{V}_\circ^\top$ is $(N - R \times N)$.
We may then express the decomposition of $\pmb{\mathcal{A}}$ as
\begin{align}
    \label{eq:A}
    \pmb{\mathcal{A}} =
    \left(
    \begin{array}{c|c}
            \mathbf{U}_\bullet & \mathbf{U}_\circ
        \end{array}
    \right)
    \left(
    \begin{array}{c|c}
            \mathbf{S}_\bullet & \mathbf{0}       \\
            \hline
            \mathbf{0}         & \mathbf{S}_\circ
        \end{array}
    \right)
    \left(
    \begin{array}{cc}
            \mathbf{V}_\bullet^\top \\
            \hline
            \mathbf{V}_\circ^\top
        \end{array}
    \right)
    \quad.
\end{align}
Inserting this into Equation~(\ref{eq:fAy}), we have
\begin{proof}{test_decomposition}
    \label{eq:ydecomp}
    \mathbf{f} - \mathbf{1} & = \pmb{\mathcal{A}} \, \mathbf{y}
    \nonumber                               \\
    & =
    \left(
    \begin{array}{c|c}
            \mathbf{U}_\bullet & \mathbf{U}_\circ
        \end{array}
    \right)
    \left(
    \begin{array}{c|c}
            \mathbf{S}_\bullet & \mathbf{0}       \\
            \hline
            \mathbf{0}         & \mathbf{S}_\circ
        \end{array}
    \right)
    \left(
    \begin{array}{cc}
            \mathbf{V}_\bullet^\top \\
            \hline
            \mathbf{V}_\circ^\top
        \end{array}
    \right) \mathbf{y}
    \nonumber                               \\[0.5em]
    & =
    \mathbf{U}_\bullet \, \mathbf{S}_\bullet \, \mathbf{V}_\bullet^\top \, \mathbf{y}
    +
    \mathbf{U}_\circ \, \mathbf{S}_\circ \, \mathbf{V}_\circ^\top \, \mathbf{y}
    \nonumber                               \\[0.5em]
    & =
    \mathbf{U}_\bullet \, \mathbf{S}_\bullet (\,\mathbf{I}\,) \mathbf{V}_\bullet^\top \, \mathbf{y}
    +
    \mathbf{U}_\circ \, \mathbf{S}_\circ (\,\mathbf{I}\,) \mathbf{V}_\circ^\top \, \mathbf{y}
    \nonumber                               \\[0.5em]
    & =
    \mathbf{U}_\bullet \, \mathbf{S}_\bullet (\mathbf{V}_\bullet^\top \mathbf{V}_\bullet) \mathbf{V}_\bullet^\top \, \mathbf{y}
    +
    \mathbf{U}_\circ \, \mathbf{S}_\circ (\mathbf{V}_\circ^\top \mathbf{V}_\circ) \mathbf{V}_\circ^\top \, \mathbf{y}
    \nonumber                               \\[0.5em]
    & =
    (\mathbf{U}_\bullet \, \mathbf{S}_\bullet \, \mathbf{V}_\bullet^\top) \mathbf{V}_\bullet \mathbf{V}_\bullet^\top \, \mathbf{y}
    +
    (\mathbf{U}_\circ \, \mathbf{S}_\circ \, \mathbf{V}_\circ^\top) \mathbf{V}_\circ \mathbf{V}_\circ^\top \, \mathbf{y}
    \nonumber                               \\[0.5em]
    & =
    (\mathbf{U}_\bullet \, \mathbf{S}_\bullet \, \mathbf{V}_\bullet^\top) \, \mathbf{y}_\bullet
    +
    (\mathbf{U}_\circ \, \mathbf{S}_\circ \, \mathbf{V}_\circ^\top) \, \mathbf{y}_\circ
\end{proof}
where we define
\begin{proof}{test_PN}
    \label{eq:yrow}
    \mathbf{y}_\bullet & \equiv \mathbf{P} \, \mathbf{y}
    \\
    \label{eq:ynull}
    \mathbf{y}_\circ   & \equiv \mathbf{N} \, \mathbf{y}
    \\[0.5em]
    \mathbf{P} & \equiv \mathbf{V}_\bullet \mathbf{V}_\bullet^\top
    \\
    \mathbf{N} & \equiv \mathbf{V}_\circ \mathbf{V}_\circ^\top
    \quad,
\end{proof}
and we used the fact that since $\mathbf{V}^\top$ is orthogonal,
\begin{proof}{test_orthogonality}
    \mathbf{V}_\bullet^\top \mathbf{V}_\bullet & = \mathbf{I}
    \nonumber                                                 \\
    \mathbf{V}_\circ^\top \mathbf{V}_\circ     & = \mathbf{I}
    \quad,
\end{proof}
where $\mathbf{I}$ is the identity matrix.
Now, recalling that $R$ is the number of nonzero singular values in
$\mathbf{S}$, it is evident from Equation~(\ref{eq:S}) that
\begin{align}
    \label{eq:S0}
    \mathbf{S}_\circ = \mathbf{0}
    \quad.
\end{align}
Therefore we may write Equation~(\ref{eq:ydecomp}) as
\begin{proof}{test_decomposition}
    \boxed{
        \mathbf{f} = \mathbf{1} + \pmb{\mathcal{A}} \, \mathbf{y}_\bullet
        +
        \mathbf{0} \, \mathbf{y}_\circ
    }
\end{proof}
where the fact that $\mathbf{U}_\bullet \, \mathbf{S}_\bullet \, \mathbf{V}_\bullet^\top = \pmb{\mathcal{A}}$
follows directly from Equations~(\ref{eq:A}) and (\ref{eq:S0}).
This is the decomposition of the surface map $\mathbf{y}$ into
its preimage $\mathbf{y}_\bullet$ and null space
$\mathbf{y}_\circ$ components.

Two things should be noted concerning this derivation. First,
because of the orthogonality of $\mathbf{V}$,
\begin{proof}{test_PN}
    \mathbf{V} \mathbf{V}^\top &= \mathbf{I}
    \nonumber                                                 \\
    \mathbf{V}_\bullet \mathbf{V}_\bullet^\top  + \mathbf{V}_\circ \mathbf{V}_\circ^\top &= \mathbf{I}
    \nonumber \\
    \mathbf{P} + \mathbf{N} &= \mathbf{I}
    \quad,
\end{proof}
so it follows from Equations~(\ref{eq:yrow}) and (\ref{eq:ynull})
that $\mathbf{y} = \mathbf{y}_\bullet + \mathbf{y}_\circ$.

Second, the singular value decomposition is not always unique. The
matrix of singular values $\mathbf{S}$ is unique provided we arrange
them in decreasing order, but if there are degenerate or zero-valued
singular values (as is the case here), the matrices $\mathbf{U}$ and
$\mathbf{V}^\top$ (and thus $\mathbf{V}^\top_\bullet$ and $\mathbf{V}^\top_\circ$)
are only well-defined up to arbitrary unitary
transformations. However, the quantities
$\mathbf{V}_\bullet\mathbf{V}^\top_\bullet$
and
$\mathbf{V}_\circ\mathbf{V}^\top_\circ$
(which define the operators $\mathbf{P}$ and $\mathbf{N}$ above) \emph{are}
unique, so the decomposition into null space and preimage is always well-defined.

\end{document}